\documentclass{aa}
\usepackage[varg]{txfonts}
\usepackage{amssymb}
\usepackage{subcaption}
\usepackage{natbib}
\usepackage{tabu}
\usepackage{placeins}

\bibpunct{(}{)}{;}{a}{}{,}

\begin{document}

\title{Characterization of the June epsilon Ophiuchids meteoroid stream and the comet 300P/Catalina}

\author{Pavol Matlovič\inst{1}
  \and Leonard Kornoš\inst{1}
  \and Martina Kováčová\inst{1}
  \and Juraj Tóth\inst{1}
  \and Javier Licandro\inst{2,}\inst{3}}

\institute{Faculty of Mathematics, Physics and Informatics,
  Comenius University, Bratislava, Slovakia\\
  \email{matlovic@fmph.uniba.sk}  
  \and Instituto de Astrofísica de Canarias (IAC), C/Vía Láctea sn, 38205 La Laguna, Spain
  \and Departamento de Astrofísica, Universidad de La Laguna, 38206 La Laguna, Tenerife, Spain}

\date{Received 2020}

\abstract {} {Prior to 2019, the June epsilon Ophiuchids (JEO) were known as a minor unconfirmed meteor shower with activity that was considered typically moderate for  bright fireballs. An unexpected bout of enhanced activity was observed in June 2019, which even raised the possibility that it was linked to the impact of the small asteroid 2019 MO near Puerto Rico. Early reports also point out the similarity of the shower to the orbit of the comet 300P/Catalina. We aim to analyze the orbits, emission spectra, and material strengths of JEO meteoroids to provide a characterization of this stream, identify its parent object, and evaluate its link to the impacting asteroid 2019 MO.} {Our analysis is based on a sample of 22 JEO meteor orbits and four emission spectra observed by the AMOS network at the Canary Islands and in Chile. The meteoroid composition was studied by spectral classification based on relative intensity ratios of Na, Mg, and Fe. Heliocentric orbits, trajectory parameters, and material strengths were determined for each meteor and the mean orbit and radiant of the stream were calculated. The link to potential parent objects was evaluated using a combination of orbital-similarity D-criteria and backwards integration of the orbit of comet 300P and the JEO stream.} {We confirm the reports of an unexpected swarm of meteoroids originating in the JEO stream. JEO meteoroids have low material strengths characteristic for fragile cometary bodies, and they exhibit signs of a porous structure. The emission spectra reveal slightly increased iron content compared to all other measured cometary streams, but they are generally consistent with a primitive chondritic composition. Further dynamical analysis suggests that the JEO stream is likely to originate from comet 300P/Catalina and that it was formed within the last 1000 years. Over longer timescales, the meteoroids in the stream move to chaotic orbits due to the turbulent orbital evolution of the comet. Our results also suggest that the impact of the small asteroid 2019 MO on June 22 was not connected to the JEO activity.} {}

\maketitle

%

\section{Introduction}  \label{introduction}

With the increasing coverage of network meteor observations around the world, we get more information about the activities of minor meteor showers. Each year, several new potential showers are added to the International Astronomical Union (IAU) database. While some of these may include duplicates of existing streams and others might not show sufficient annual activity to be confirmed, they are generally likely to include real new meteoroid streams from a yet-to-be confirmed parent asteroid or comet. Studies of such streams therefore provide valuable information about the structure of the meteoroid complex in the near-Earth space and offer evidence of past activities of their parent objects.

June epsilon Ophiuchids (JEO, \#459) are a minor shower first reported by \citet{2014me13.conf..217R}. Only few meteors were assigned to the proposed shower, with a few more observations reported from CAMS data \citep{2016Icar..266..331J}. The early results pointed out a Jupiter-family type orbit similar to comet 300P/Catalina which, due to its low activity, was originally identified as the asteroid 2005 JQ5. 

In 2019, the shower exhibited an unexpected bout of significantly increased activity, including numerous reported fireballs. Interestingly, the time of the shower outburst and its entry speed correlates with the impact of a small asteroid 2019 MO observed on June 22 off the southern coast of Puerto Rico. The asteroid, measuring a few meters in size, was originally discovered by the ATLAS project\footnote{https://atlas.fallingstar.com/home.php}. To date, it is only the fourth asteroid to be discovered prior to impact.

The 2019 outburst of JEO was first noted by Peter Jenniskens (electronic telegram CBET 4642) and subsequent observations were reported by multiple networks around the world. A preliminary analysis of these observations, combined with a search in the EDMOND database \citep{2019eMetN...4..201R}, have confirmed an unusual increase in JEO activity in 2019. It was noted at the time that the link to the impacting asteroid 2019 MO based on orbital similarity is ambiguous and may be just a result of the densely populated orbital parameter space in this region.

A more detailed analysis of the stream orbit and physical properties its meteoroids is needed to confirm or refute the link to 2019 MO and 300P/Catalina (further referred to as 300P). Here we present an analysis of JEO orbits, emission spectra, and material strength properties, all aimed at resolving questions about the origins of this stream, which can be now considered the source of an established meteor shower. Furthermore, we use dynamical integrations to study the stability of the orbit of 300P and its modeled meteoroid stream.

\section{Observations and methods}  \label{Methods}

\begin{figure}
\centerline{\includegraphics[width=\columnwidth,angle=0]{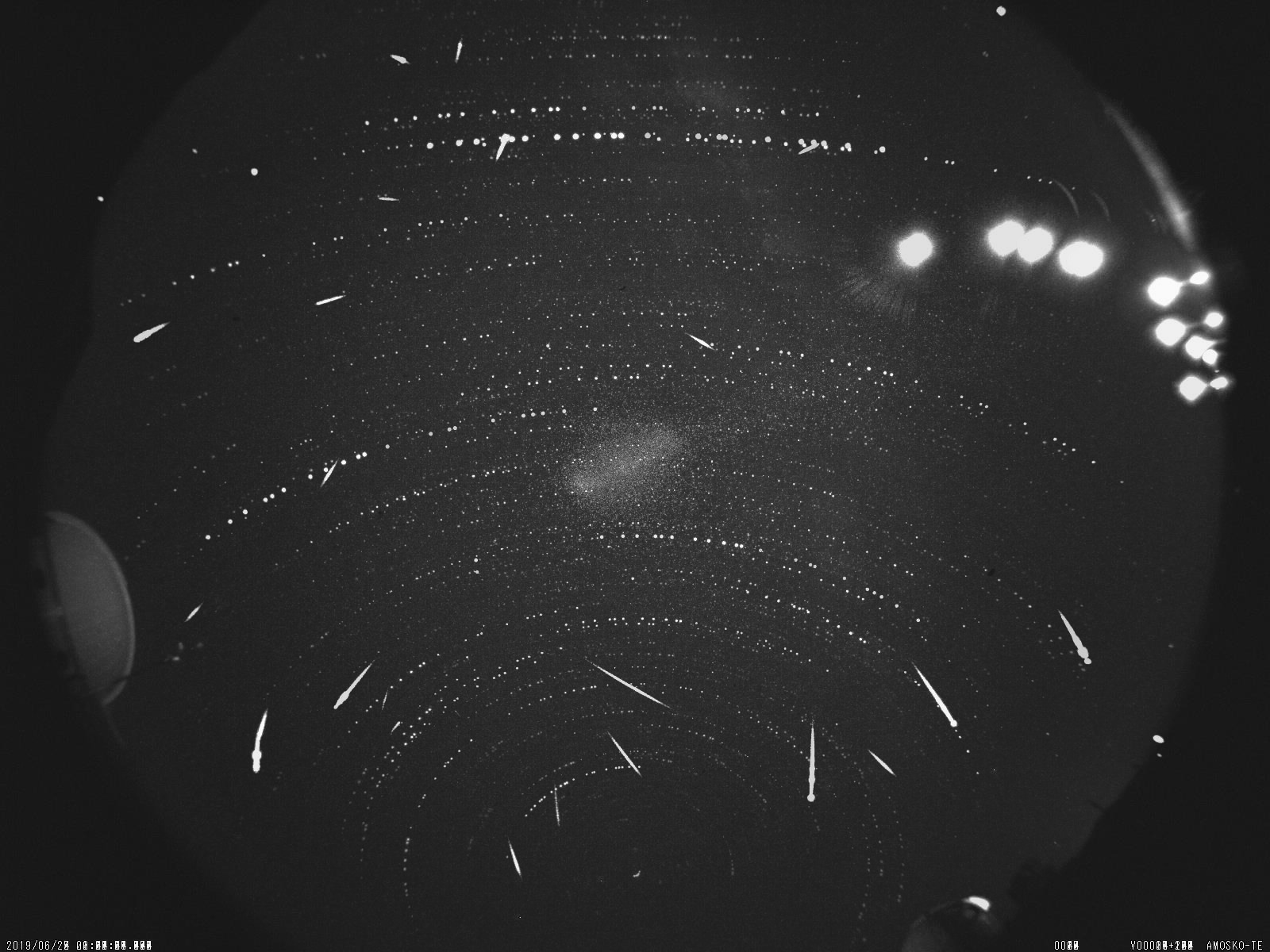}}
\caption[f1]{Composite image of JEO meteors observed between June 22-25 by the AMOS station on Teide Observatory, Canary Islands.}
\label{composite}
\end{figure}

We selected 24 JEO meteors observed between June 22-25 by the All-sky Meteor Orbit System (AMOS) network on the Canary Islands (Roque de los Muchachos observatory on La Palma [LP] and Teide observatory on Tenerife [TE]), and in Chile (SPACE observatory in San Pedro de Atacama [SP]). Due to bad weather conditions, no observations were obtained from the AMOS systems in Slovakia. Other systems in Hawaii and Chile captured some activity from single stations, which are not used in this work. Two single-station spectral observations were used -- one from TE and one from SP. The assignment of these cases to the JEO stream was based on the estimated radiant position, angular speed, and measured emission spectrum consistent with confirmed JEO meteors. A composite image of JEO meteors observed at the AMOS-TE station is presented in Fig. \ref{composite}.

The standard AMOS system consists of four major components: a fish-eye lens, an image intensifier, a projection lens, and a digital video camera. The resulting field of view (FOV) of AMOS is 180$^{\circ}$ x 140$^{\circ}$ with an image resolution of 1600x1200 pixels (8.4 arcmin/pixel), and a video frame rate of 20 fps. Limiting magnitude for stars is about +5 mag for a single frame, and the detection efficiency is lower for moving objects, approximately +4 mag at typical meteor speeds due to the trailing loss. More information about the global AMOS network is available in \citet{2015P&SS..118..102T} and \citet{2013pimo.conf...18Z}.

The spectra analyzed in this work were captured by the updated version of the original AMOS-Spec system \citep{2016P&SS..123...25R, 2019A&A...629A..71M}, known as the AMOS-Spec-HR. This system is based on 2048x1536 px Point-Grey camera, 6 mm f/1.4 lens, and 1000 grooves/mm holographic grating, resulting in a 60x45$^{\circ}$ rectangular FOV and dispersion of 0.5 nm/px. The spectral resolution of the AMOS-Spec-HR is R $\approx$ 500. The limiting magnitude of the system is around -1.5 mag for spectral events and obtained recordings have frame rate of 15 fps.

Each analyzed spectrum was manually scanned in individual frames, calibrated, and fitted according to the procedure described in \citet{2019A&A...629A..71M}. After the reduction of the continuum radiation and atmospheric emission lines and bands, which are generally very faint in spectra of slow meteors, all present meteor emission lines were identified and measured. Intensity ratios of the main spectral multiplets of Na I - 1, Mg I - 2 and Fe I - 15 measured in the fitted spectra were used for the spectral classification of JEO meteoroids, in accordance with the procedures of \citet{2005Icar..174...15B}. The typical FWHM of the fitted spectral lines was between 0.9 and 1.4 nm.

Multi-station observations of 22 JEO meteoroids were used to determine their orbital and trajectory parameters. The \textit{UFOCapture} detection software \citep{2009JIMO...37...55S} was used during real-time observations to detect and record all of the studied meteors. Star and meteor coordinates from each frame of each event were measured using the original \textit{AMOS} software. Meteor photometry, astrometry and orbit determination was performed using the \textit{Meteor Trajectory} (\textit{MT}) software \citep{2015pimo.conf..101K,KornosIMC17} based on the procedure of \citet{1987BAICz..38..222C} and \citet{1995A&AS..112..173B}. The precision of the AMOS astrometry is on the order of 0.02 - 0.03$^{\circ}$, which translates to an accuracy of tens of meters for atmospheric meteor trajectory.

The dynamical evolution of the proposed parent comet and JEO meteoroids was studied by backwards orbital integration using the IAS15 integrator included in the REBOUND software package. IAS15 is a non-symplectic 15th-order integrator with adaptive step-size control \citep{2015MNRAS.446.1424R}. In our integration, we took into account the Sun, planets, the Moon, and the four largest asteroids from the main belt: Ceres, Pallas, Vesta, and Hygiea. Initial positions and velocities were set from the JPL NASA Horizons system. The comet 300P, as well as the modeled JEO meteoroids that were initially evenly spread along the calculated mean JEO orbit, was represented by massless test particles.

\section{Trajectories and orbits}  \label{Orbits}

All meteors analyzed here were observed across three nights between June 22 and June 25 2019 (90.94$^{\circ}$ < $\lambda_{\odot}$ < 92.98$^{\circ}$). Trajectory parameters and orbits of 22 JEO meteoroids determined from multi-station observation are given in Tables \ref{physical} and \ref{orbits}. Our sample includes meteors of +1.3 to -5.4 absolute magnitude, with four JEO meteors brighter than -4 mag, which can be classified as fireballs. Additionally, two JEO fireballs were observed as single-station events captured with the spectrum. It was reported that the CAMS network observed around 88 JEO orbits during the 2019 outburst \citep{2019eMetN...4..201R}. Such general high activity within the stream and in the production of fireballs was not previously observed or anticipated, which points to the activity of the parent object of the stream. 

\begin{figure}[!t]
    \centering
    \begin{subfigure}[b]{0.5\columnwidth}
      \centering
      \includegraphics[width=\textwidth]{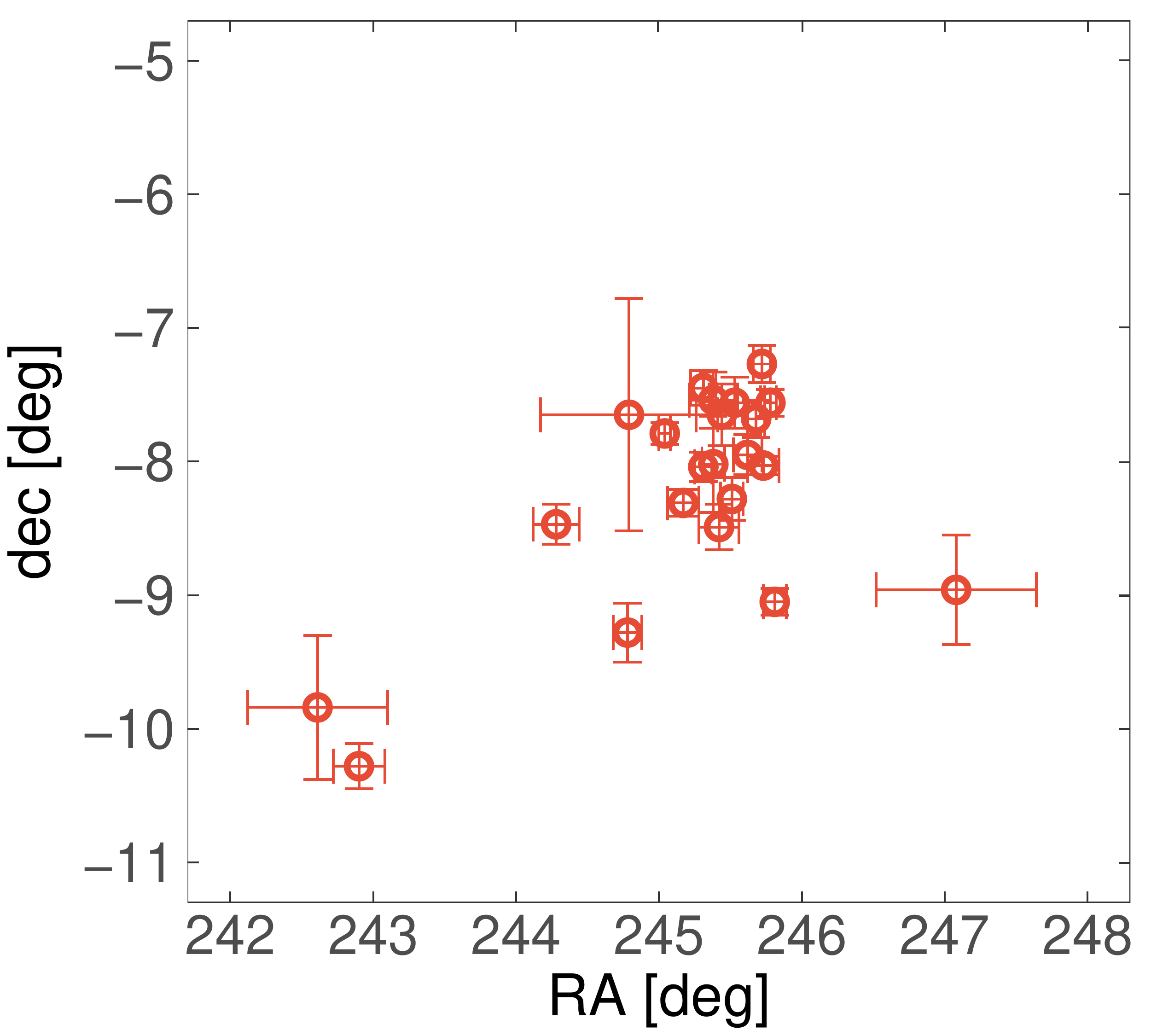}    
      \label{fig:1}
    \end{subfigure}%
    ~
    \begin{subfigure}[b]{0.5\columnwidth}
      \centering
      \includegraphics[width=\textwidth]{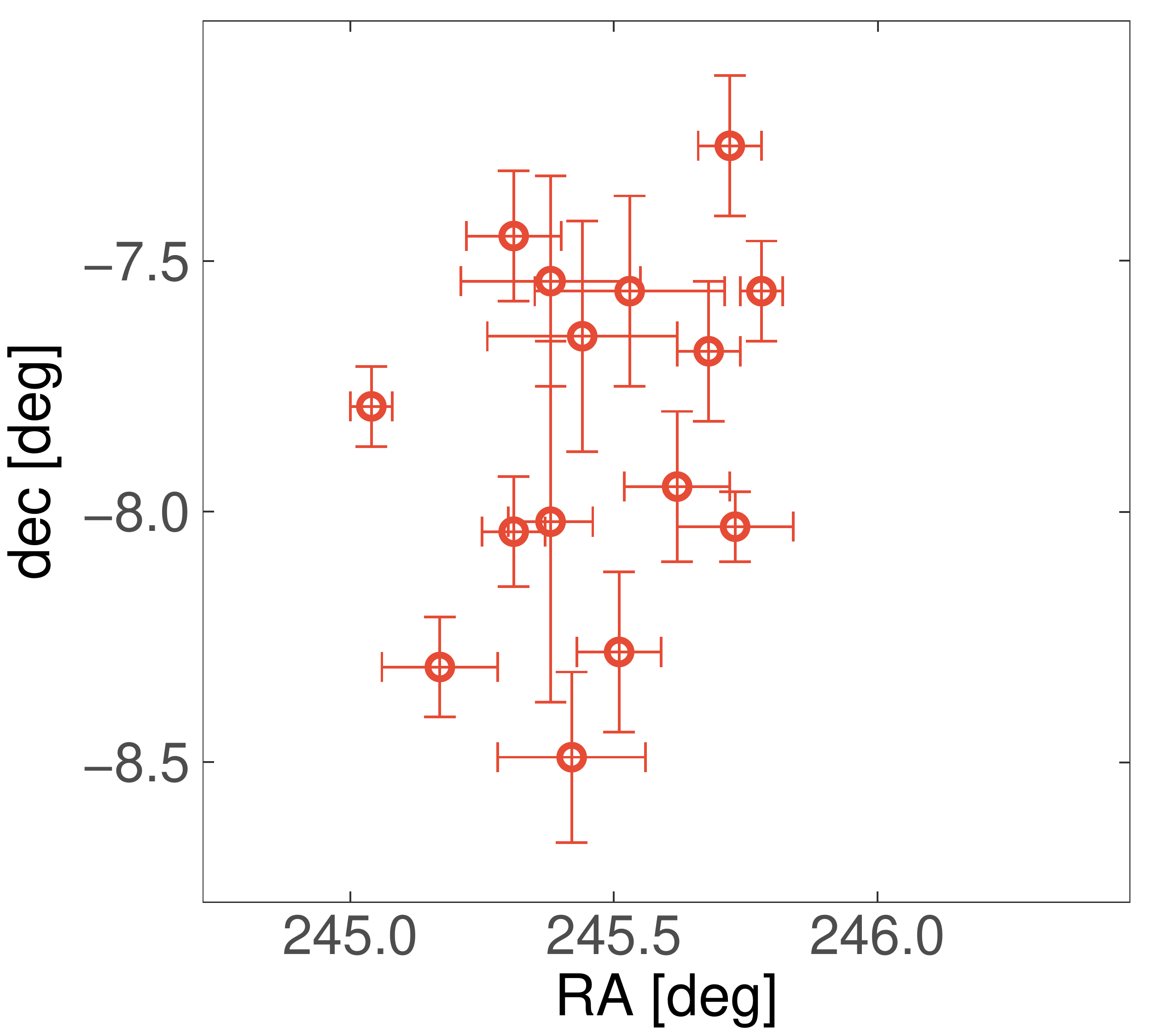}      
      \label{fig:2}
    \end{subfigure}%
    \caption{Geocentric radiant distribution of all 22 JEO meteors observed during the increased activity between June 22 and June 25 2019 (left) and zoom-in at the core of the shower outburst (right).}
    \label{radiant}
  \end{figure}

The radiant distribution of JEO meteors (Fig. \ref{radiant}) shows a slightly dispersed radiant area with the core of the outburst at 245$^{\circ}$ < RA$_g$ < 246$^{\circ}$ and -8.5$^{\circ}$ < dec$_g$ < -7$^{\circ}$. The mean radiant position is shifted slightly from the previous estimates based on the limited data collected before the 2019 outburst (Table \ref{meanorbit}). 

\begin{figure}
\centerline{\includegraphics[width=0.8\columnwidth,angle=0]{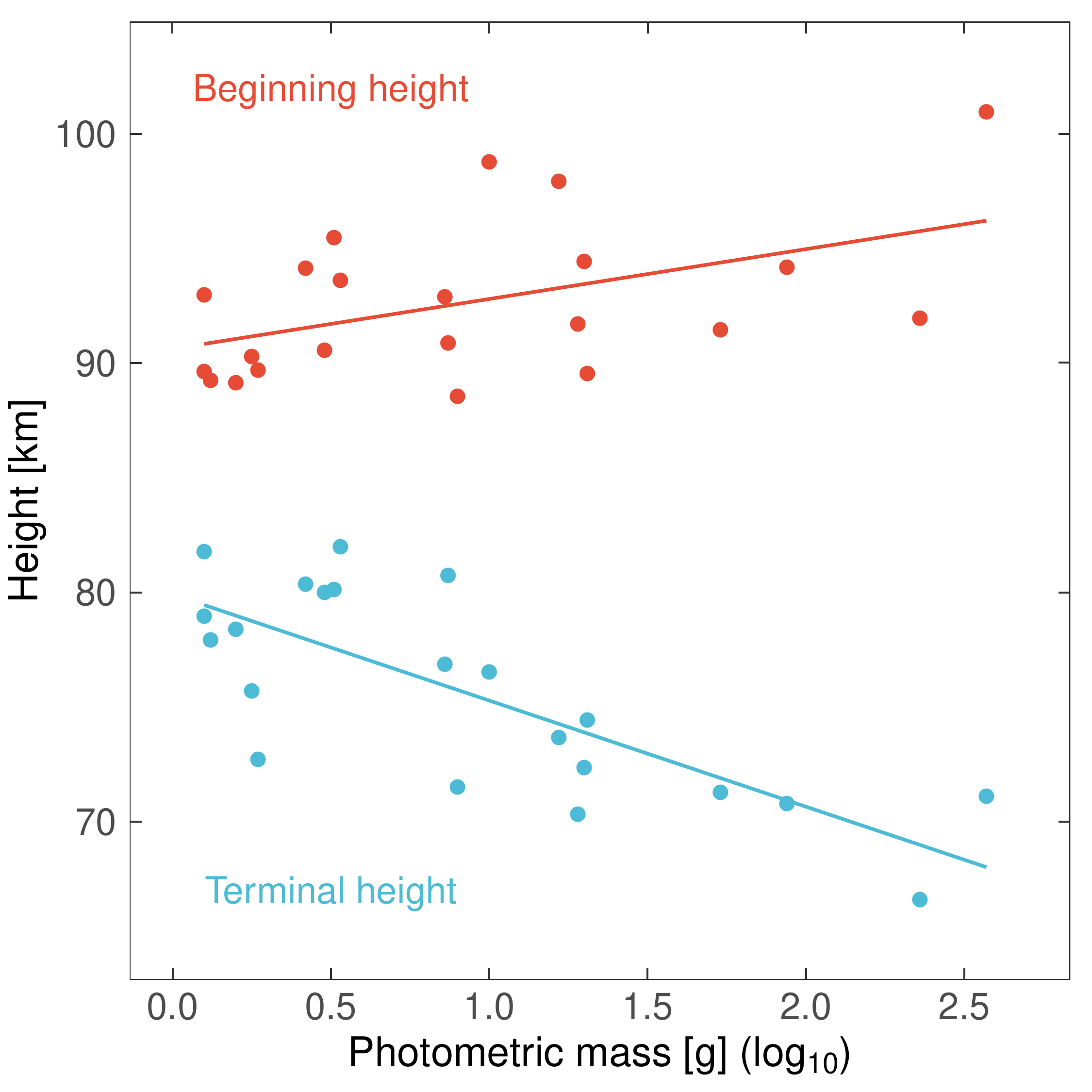}}
\caption[f1]{Meteor beginning and terminal heights as functions of the meteoroid photometric mass.}
\label{HbHe}
\end{figure}

\begin{figure}
\centerline{\includegraphics[width=0.95\columnwidth,angle=0]{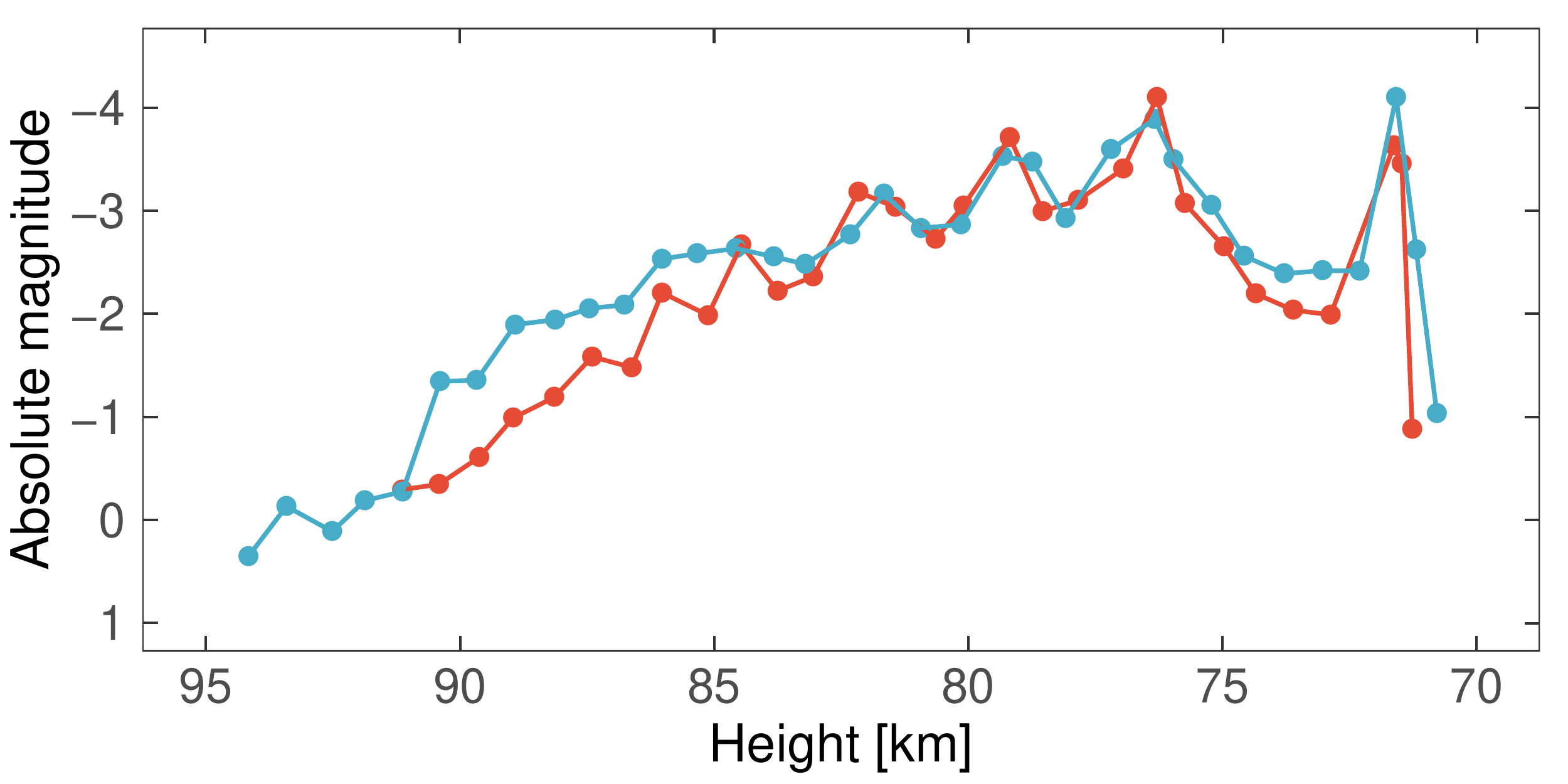}}
\caption[f1]{Light curve of a bright JEO meteor (M20190624\_222248) as observed from two AMOS stations on La Palma (red) and Tenerife (blue).}
\label{lightcurve}
\end{figure}

\begin{table*}[]
\centering
\small\begin{center}
\caption {Mean orbit of JEO meteoroid stream $\pm$ the standard deviation and a comparison to a mean orbit determined from 2019 CAMS data \citep{2019eMetN...4..201R} and the current IAU MDC orbit from data collected before 2019 \citep{2016Icar..266..331J}.} 
\vspace{0.1cm}
\resizebox{\textwidth}{!}{\begin{tabular}{llllllllll}
\hline\hline\\
\multicolumn{1}{c}{}& %
\multicolumn{1}{c}{$RA_g$} & %
\multicolumn{1}{c}{$dec_g$} & %
\multicolumn{1}{c}{$v_{g}$} & %
\multicolumn{1}{c}{$a$} & %
\multicolumn{1}{c}{$e$} & %
\multicolumn{1}{c}{$q$} & %
\multicolumn{1}{c}{$i$} & %
\multicolumn{1}{c}{$\omega$}& %
\multicolumn{1}{c}{$\Omega$} %
\\
\hline\\
This work & 245.19 $\pm$ 0.95 & -8.23 $\pm$ 0.80 & 14.22 $\pm$ 0.48 & 2.79 $\pm$ 0.20 & 0.68 $\pm$ 0.02 & 0.884 $\pm$ 0.006 & 5.16 $\pm$ 0.42 & 227.37 $\pm$ 0.95 & 91.68 $\pm$ 0.59 \vspace{0.1cm}\\

CAMS 2019 & 245.20 $\pm$ 1.30  & -7.40 $\pm$ 2.00 & 14.20 $\pm$ 1.10 & 2.69 $\pm$ 0.52 & 0.67            & 0.885 $\pm$ 0.011 & 5.30 $\pm$ 0.90 & 227.30 $\pm$ 1.90 & 92.20 $\pm$ 1.10  \vspace{0.1cm}\\

IAU MDC   & 244.70             & -8.80             & 14.90             & 2.53            & 0.66           & 0.866             & 4.90             & 230.30             & 89.10            \vspace{0.1cm} \\
\hline
\end{tabular}}
\label{meanorbit}
\end{center}
\end{table*}

Table \ref{meanorbit} also shows the mean geocentric speed and orbit of JEO meteoroids and a comparison to the results from 2019 CAMS data and the IAU MDC orbit from pre-2019 data. Our results are consistent with the 2019 CAMS data \citep{2019eMetN...4..201R} and differ (even considering the standard deviation of our results) from the previously estimated parameters \citep{2016Icar..266..331J}, most notably in the geocentric velocity (by 0.7 km\,s\textsuperscript{-1}), semi-major axis (by 0.25 au), and inclination (by 0.26$^{\circ}$). The stream has a Jupiter-family type orbit ($T_J =$ 2.93 $\pm$ 0.13), suggesting a short-period parent object.

The beginning heights of JEO meteoroids ranged between 88 and 101 km and generally increased with the original meteoroid mass (Fig. \ref{HbHe}). The terminal heights ranged from 82 to 66 km and decreased with mass. This behavior is not unusual and has been observed among other meteoroid streams (e.g., \citet{2004A&A...428..683K}, \citet{2019A&A...629A.137S}). With the exception of the second-largest body in our sample ($\approx$ 230 g), all JEO meteoroids were disrupted completely before reaching the 70 km altitude, suggesting fragile cometary structure (further discussed in Section \ref{Structure}). 

Light curves of brighter JEO meteors usually show multiple flares of brightness, typically exhibiting a final bright flare shortly before the end of the luminous trail (Fig. \ref{lightcurve}). The observed flares suggest a porous structure for JEO meteoroids and explosive gross disruption typically at 75 -- 70 km heights. Flares of brightness were not recognizable among fainter JEO meteors.

\section{Spectra and material properties}  \label{Structure}

\begin{figure}
\centerline{\includegraphics[width=0.96\columnwidth,angle=0]{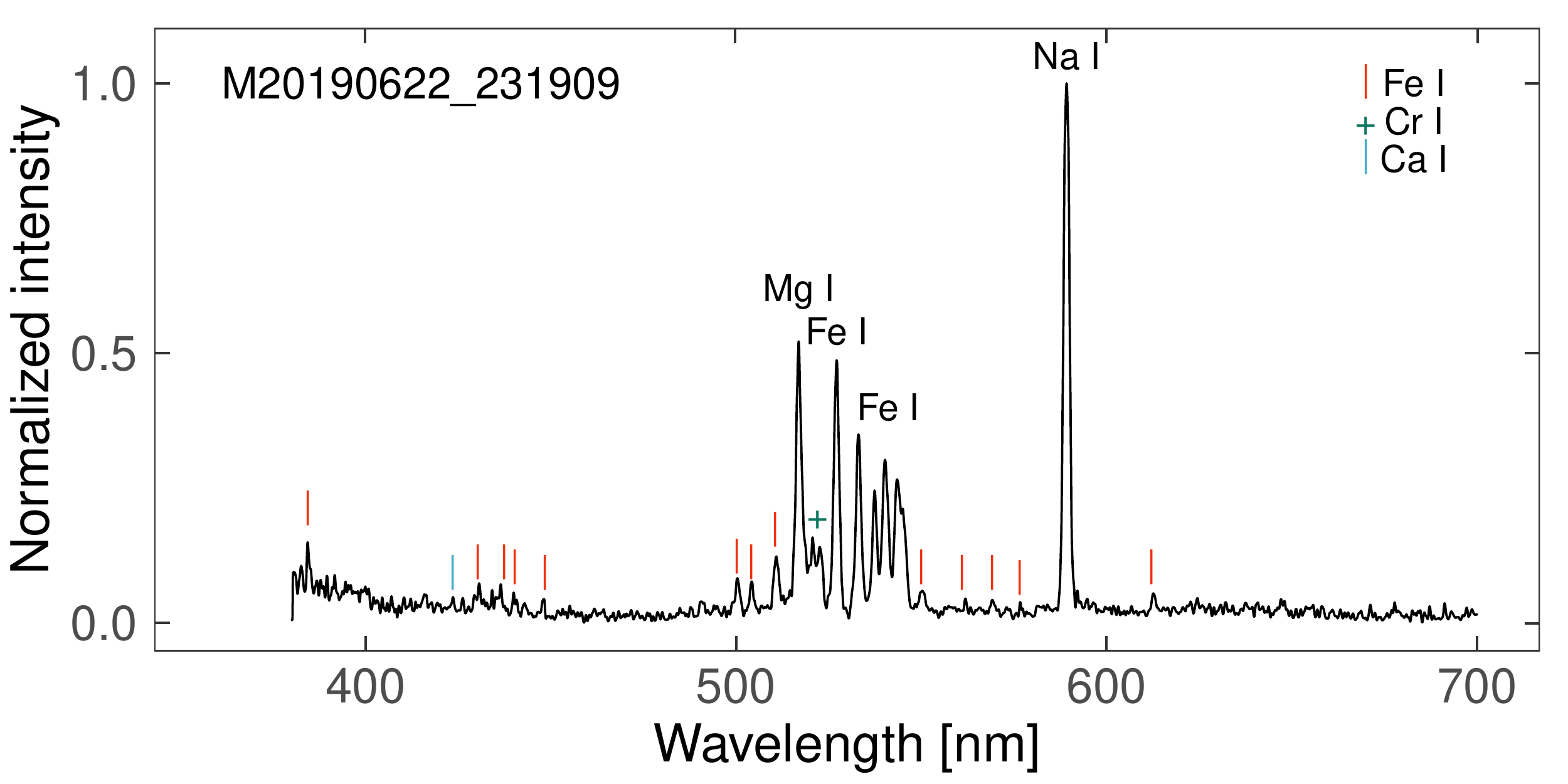}}
\centerline{\includegraphics[width=0.96\columnwidth,angle=0]{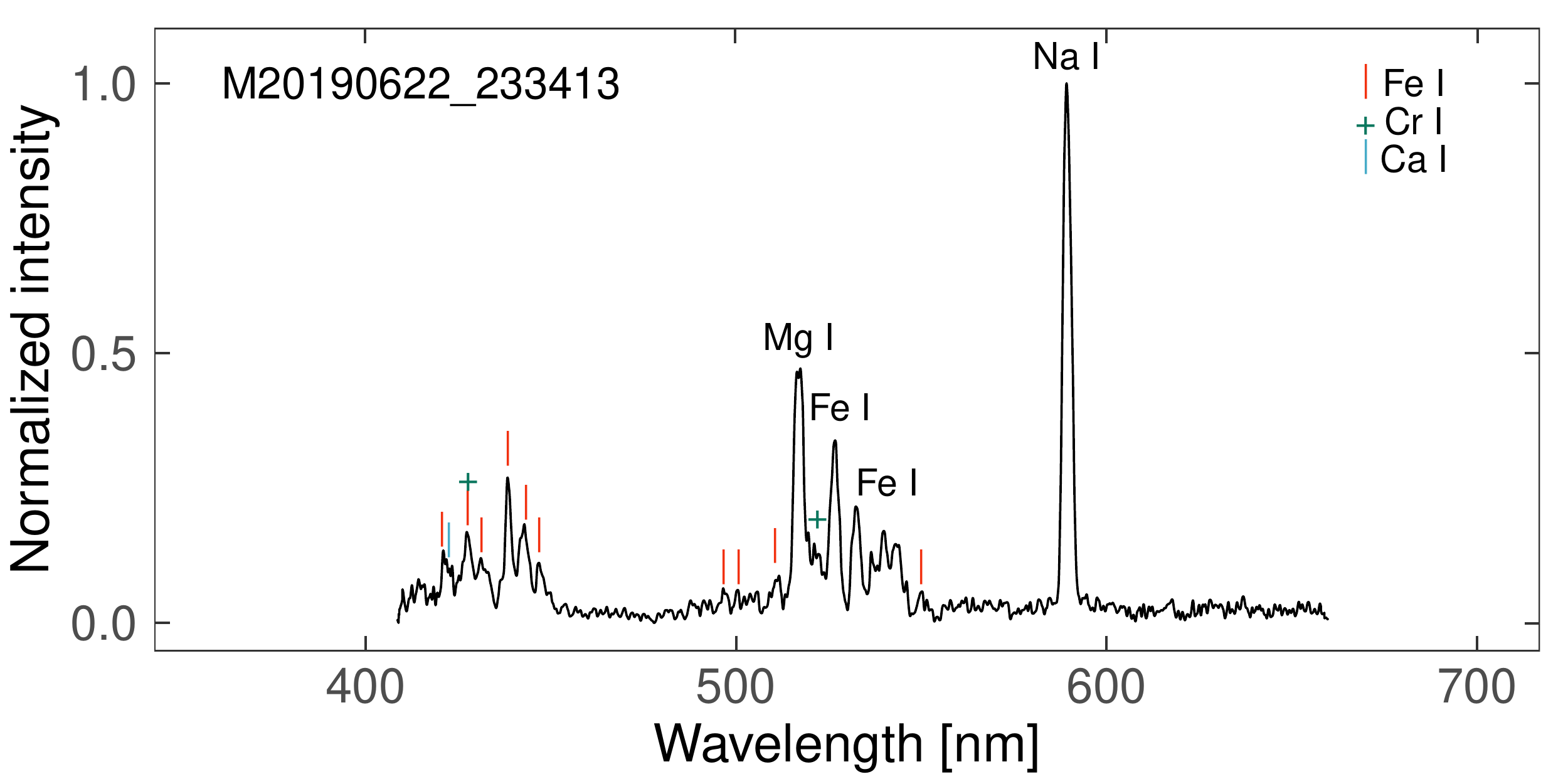}}
\centerline{\includegraphics[width=0.96\columnwidth,angle=0]{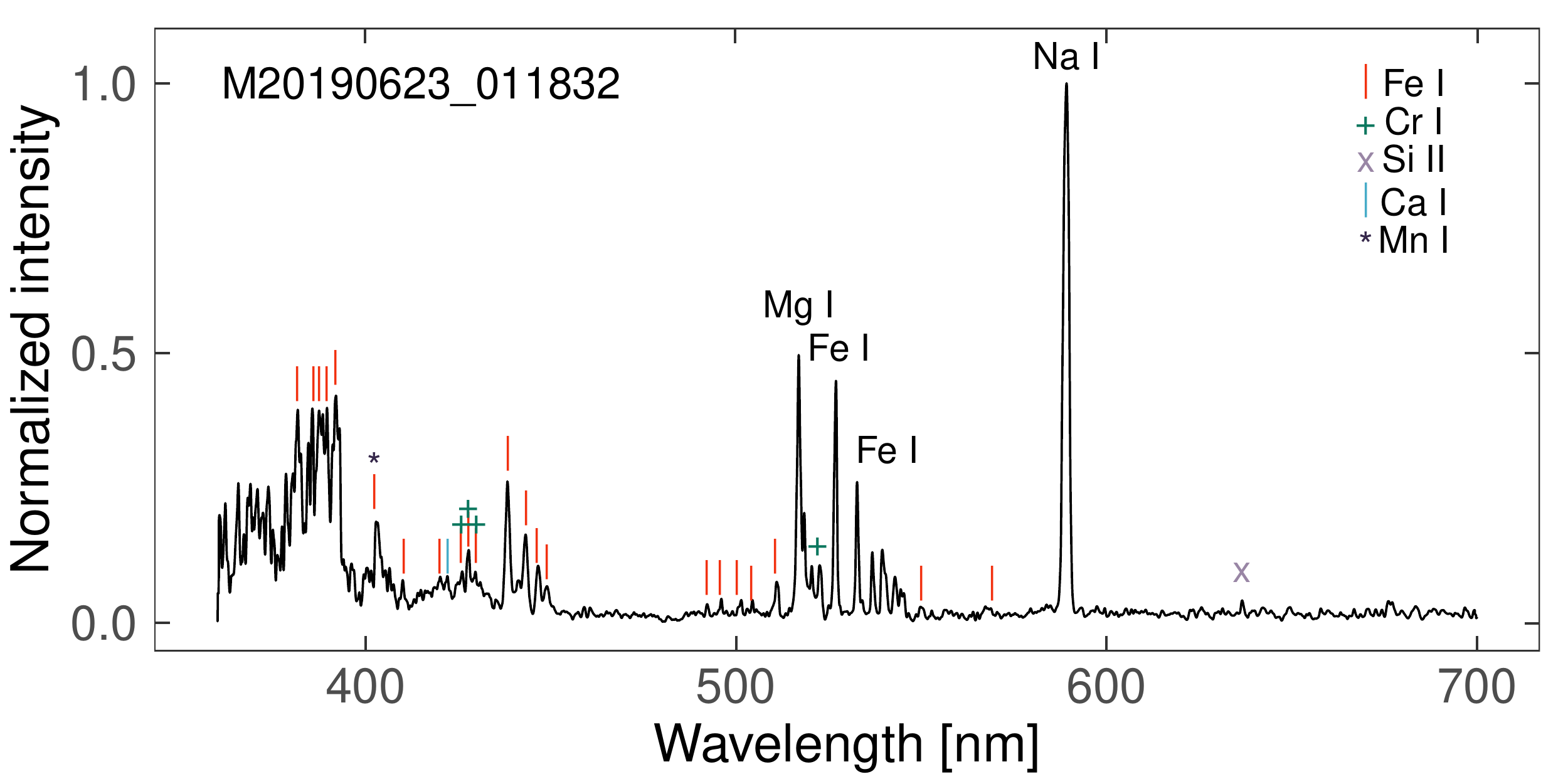}}
\centerline{\includegraphics[width=0.96\columnwidth,angle=0]{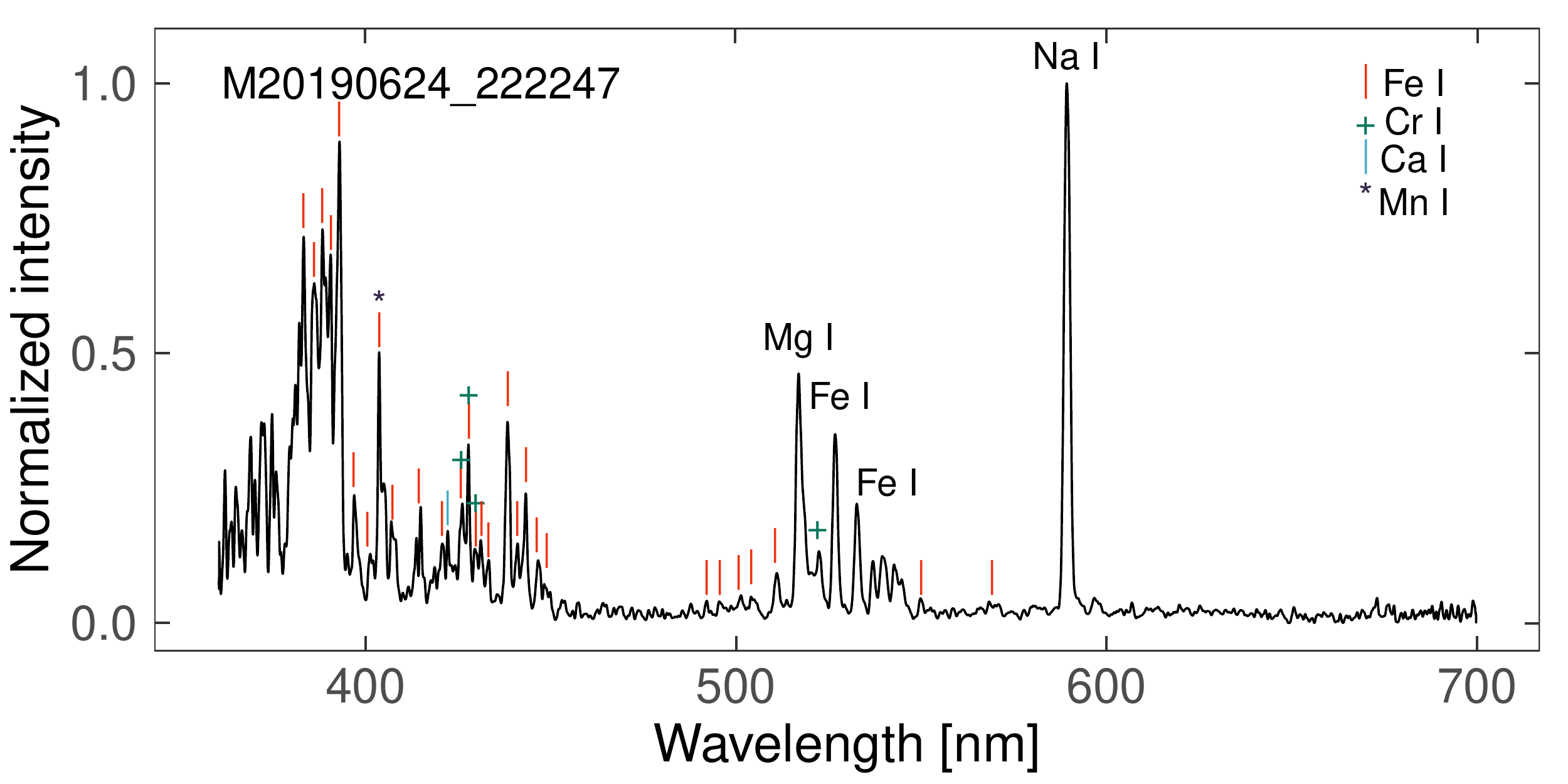}}
\caption[f1]{Visible spectra of four JEO meteors with all identified emission lines. Major multiplets used for the spectral classification (Mg I - 2, Na I - 1 and Fe I - 15) are highlighted. Lines at wavelengths below 380 nm belong dominantly to Fe and lay in region with low spectral sensitivity of the system.}
\label{spectra}
\end{figure}

Four emission spectra were observed by the global AMOS network during June 22 -- June 25 2019, all belonging to the JEO stream. The normalized spectral profiles in visible range are displayed in Fig. \ref{spectra}. All four meteors exhibit similar spectral features with corresponding Na/Mg/Fe intensity ratios in the 500 -- 600 nm region. This part of the spectrum lays in the region of high spectral sensitivity of the AMOS-Spec-HR \citep{2019A&A...629A..71M} and is typically used to characterize the spectral type of a meteoroid \citep{2005Icar..174...15B}. The 500 -- 600 nm region was fitted by all contributing emission lines taken from published line lists \citep{1994A&AS..103...83B} and the NIST database \citep{NIST_ASD}. The intensities of the Na I - 1, Mg I - 2 and Fe I - 15 multiplets were measured and their ratios were used for the spectral classification.

\begin{figure}
\centerline{\includegraphics[width=\columnwidth,angle=0]{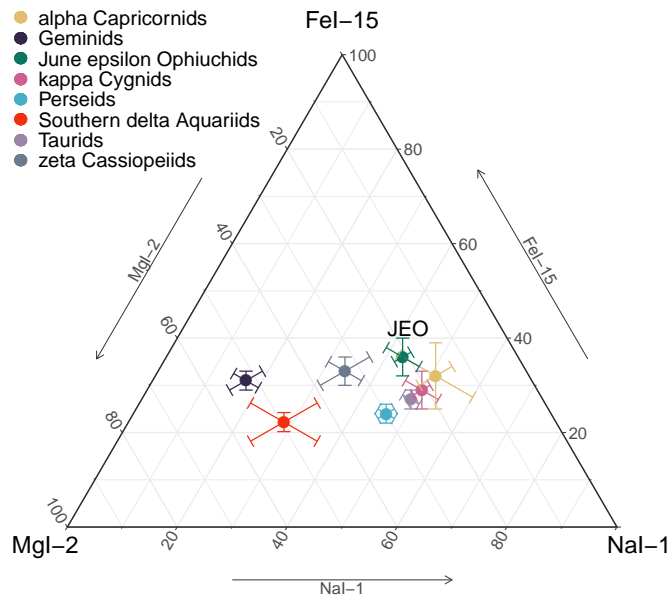}} \caption[f1]{Mean spectral classification (mean Na I/Mg I/Fe I intensity ratios and standard error of the mean) of JEO meteoroids and comparison to other major meteoroids streams from \citet{2019A&A...629A..71M}.}
\label{ternaryAVR}
\end{figure}

The mean spectral classification of JEO meteoroids and comparison with other major meteoroid streams is given in Fig. \ref{ternaryAVR}. The spectra do not significantly differ from other cometary meteoroid streams, but show slight enhancement of iron compared to all other measured streams \citep{2019A&A...629A..71M}. 

Signs of the optical thickness of the radiating plasma, which can be responsible for the apparent increase of Fe intensity, were observed to a minimal extent only in meteor M20190622\_231909. Therefore, measurements from this case were accounted to the mean classification by lower weight. The slightly higher Fe content in JEO spectra may be simply a result of a natural heterogeneity between cometary streams. This effect likely also reflects the compositional differences between the corresponding parent comets. Spectral data for the comet 300P, which we identify as the parent object of this stream, are not available.

\begin{figure}
\centerline{\includegraphics[width=0.95\columnwidth,angle=0]{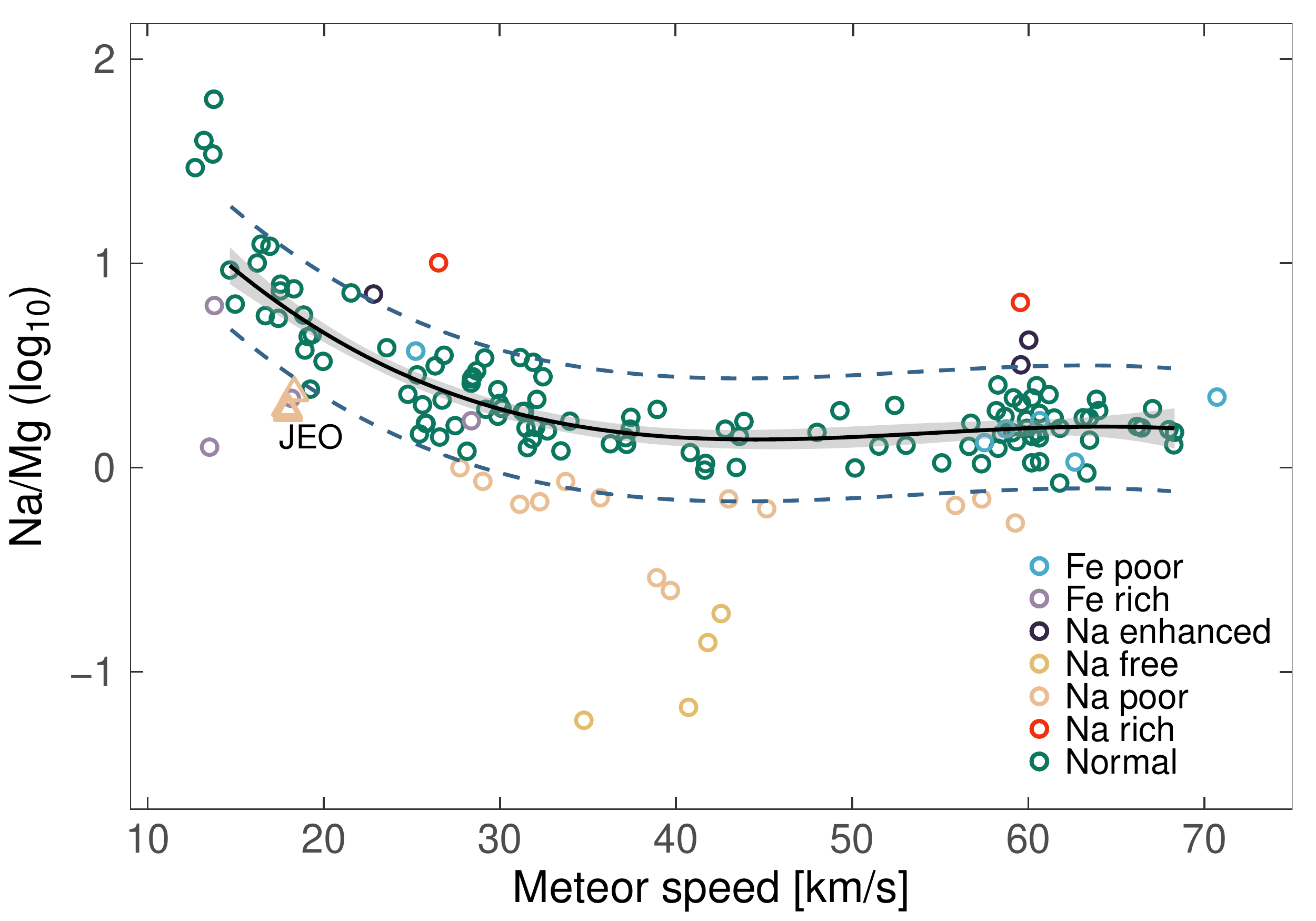}} \caption[f1]{Dependency of the measured Na/Mg intensity ratio on meteor speed adapted from \citet{2019A&A...629A..71M} showing detected sodium depletion in JEO meteoroids (triangles). For the two JEO spectra with no speed information, an average JEO speed was assumed ($v_i$ = 18.0 km\,s\textsuperscript{-1}). Errors of individual JEO speeds and Na/Mg intensity ratios are within the size of their marks.}
\label{speedJEO}
\end{figure}

Furthermore, while the JEO spectra show apparently strong Na line (Fig. \ref{spectra}), the Na intensity is, due to the low excitation of Na, generally dependent on entry speed for meteors slower than 35 -- 40 km\,s\textsuperscript{-1} \citep{2005Icar..174...15B}. In fact, considering the low speed of JEO meteors ($v_i \approx$ 18.0 km\,s\textsuperscript{-1}), the spectra show slight Na depletion when compared to a large dataset of meteor spectra from \citet{2019A&A...629A..71M} (Fig. \ref{speedJEO}) and, thus, they were classified as Na-poor. The determined mean Na/Mg and Fe/Mg intensity ratios in JEO meteors are presented in Table \ref{meanPhysical}.

Sodium depletion in meteor spectra has previously been linked with processes of space weathering: thermal desorption during close perihelion approaches (q $\leq$ 0.2 au) and cosmic ray irradiation for meteors on Halley-type orbits \citep{2005Icar..174...15B, 2006A&A...453L..17K, 2019A&A...629A..71M}. With the relatively short-period, Jupiter-family type orbit of the stream (Table \ref{meanorbit}) and q $\approx$ 0.89 au, these processes do not explain the obtained Na/Mg ratios. Based on the dynamical integration discussed further in Section \ref{evolution}, we estimate that the JEO stream was formed within the last 1000 years and was not exposed to strong solar heating. It is possible that the Na depletion occurred as a result of  space weathering on the original cometary surface during its dynamical or physical evolution.

Other lines identified in JEO spectra include Cr I, Ca I, and Mn I, which are often blended with the surrounding Fe I lines (Fig. \ref{spectra}). These elements are commonly detected in meteor spectra \citep{1998SSRv...84..327C}. No atypical spectral lines were found. We did not identify any atmospheric emission lines or bands in the studied spectra. This is a result  attributed to the low speed of JEO meteors, whereupon high-temperature spectral component is not typically observed \citep{1994P&SS...42..145B}. Overall, the spectral profiles are consistent with chondritic composition found in primitive bodies. The detected slight Fe enhancement and Na depletion, however, form specific spectral signature 

The analyzed spectra differ most in the blue and near-UV region. The intensity in this region is affected by the meteor brightness and atmospheric conditions. Furthermore, the spectral sensitivity of the system below 400 nm is too low for reliable comparison of the case-to-case line intensities. However, based on the line identification, all four spectra show consistent features in this region, mostly dominated by lines of Fe I blended with the Mg I - 3 multiplet near 383.7 nm.

The material strengths of JEO meteoroids were investigated based on the empirical parameters $K_{B}$ and $P_{E}$ developed by \citet{1968SAOSR.279.....C} and \citet{1976JGR....81.6257C}. These parameters can be used to classify meteoroid material type based on determined trajectory properties \citep{1988BAICz..39..221C}. The basic classification of meteoroid material type based on $K_B$ and $P_E$ is summarized in Table \ref{KBPEclass}. Originally $K_{B}$ was developed for the classification of fainter meteors, while $P_{E}$ is more suitable for fireballs. Since our sample includes wide range of meteor magnitudes (1.3 to -5.4), we decided to classify each case by both parameters and compare the results to get overall view of their characteristic material.

\begin{table}[]
\small\begin{center}\caption {Determined mean spectral and material parameters of the JEO stream.} 
\begin{tabular}{rccccrc}
\hline\hline\\
\multicolumn{1}{l}{}& %
\multicolumn{1}{c}{Na/Mg}& %
\multicolumn{1}{c}{Fe/Mg}& %
\multicolumn{1}{c}{$K_B$}& %
\multicolumn{1}{c}{type}& %
\multicolumn{1}{c}{$P_E$}& %
\multicolumn{1}{c}{type} \\
\hline\\
JEO mean & 2.02 & 1.96 & 6.97 & C1 & -5.83 & IIIB \\ 
 $\pm$   & 0.21 & 0.79 & 0.28 &    & 0.12  &      \\ 
\hline
\end{tabular}
\label{meanPhysical}
\end{center}
\end{table}

\begin{table}[]
\small\begin{center}\caption {Meteoroid material strength classification based on parameters $K_B$ and $P_E$ following \citet{1988BAICz..39..221C}.} 
\begin{tabular}{lcc}
\hline\hline\\
\multicolumn{1}{l}{Material type}& %
\multicolumn{1}{c}{$K_B$ type}& %
\multicolumn{1}{c}{$P_E$ type} \\
\hline\\
Fragile cometary / Draconid-type & D & IIIB \vspace{0.2cm} \\
Regular cometary  & C & IIIA \vspace{0.2cm} \\
Dense cometary  & B & - \vspace{0.2cm} \\
Carbonaceous    & A & II \vspace{0.2cm} \\
Ordinary chondrite - asteroids & ast & I \vspace{0.1cm}\\
\hline
\end{tabular}
\label{KBPEclass}
\end{center}
\end{table}

\begin{figure*}
    \centering
    \begin{subfigure}[b]{0.5\textwidth}
      \centering
      \includegraphics[width=\textwidth]{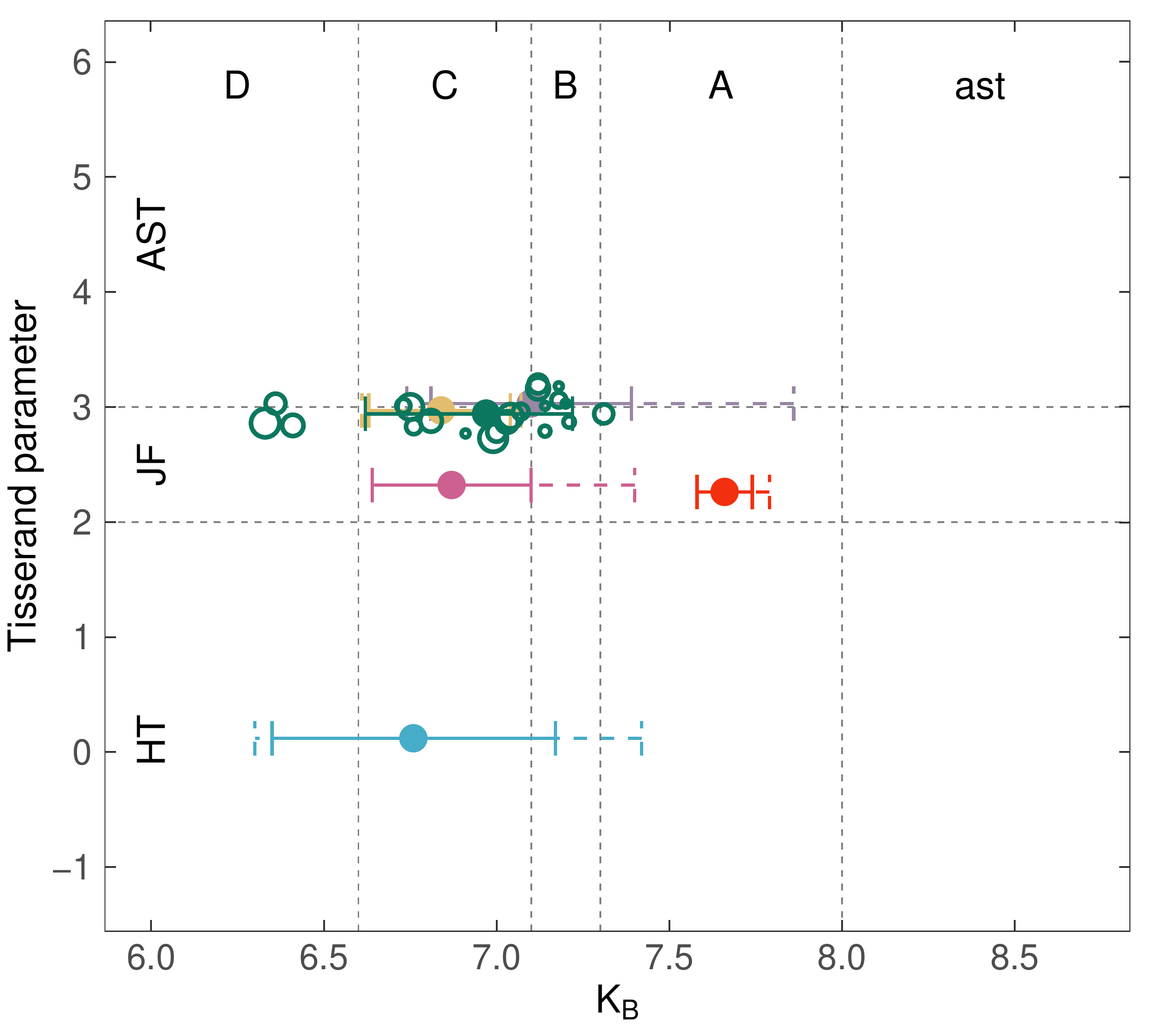}    
      \label{fig:1}
    \end{subfigure}%
    ~
    \begin{subfigure}[b]{0.5\textwidth}
      \centering
      \includegraphics[width=\textwidth]{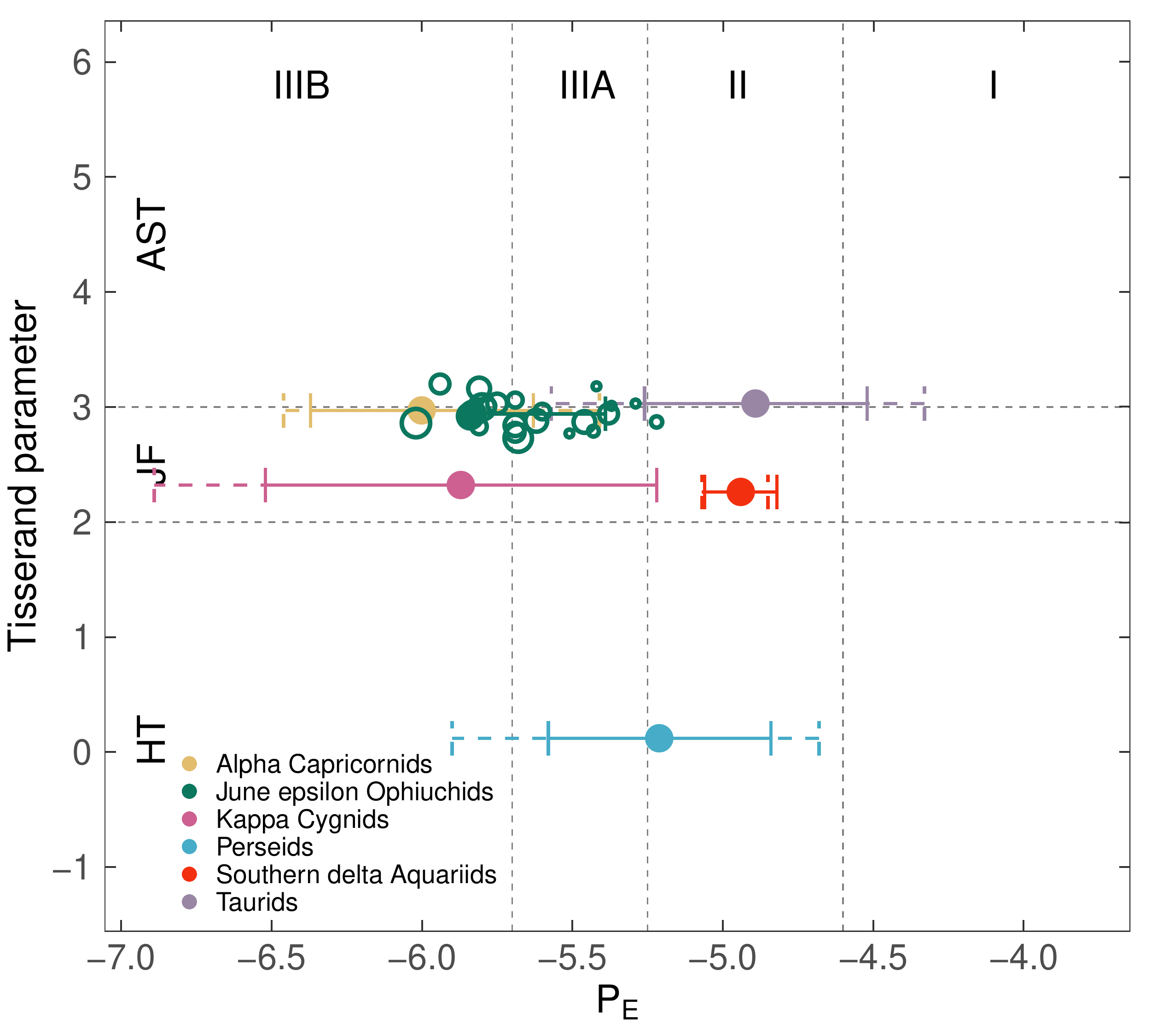}      
      \label{fig:2}
    \end{subfigure}%
    \caption{Material strength classification of the JEO meteoroids (open circles) based on the $K_B$ (left) and $P_E$ (right) parameter as a function of the Tisserand parameter relative to Jupiter. Mean $K_B$ and $P_E$ material strengths and their standard deviation (solid error bar) and overall dispersion (dashed error bar) for different meteoroid streams from \citet{2019A&A...629A..71M} are displayed for comparison. Sizes of meteoroid marks reflect relative meteoroid masses in logarithmic scale.}
    \label{KbPe}
  \end{figure*}

The obtained values are for both $K_{B}$ and $P_{E}$ characteristic for standard cometary meteoroids (Fig. \ref{KbPe}, Table \ref{physical}). Considering all meteors in our sample, JEO are, on average, classified as type C1/IIIA, but they also include several cases of the most fragile cometary material (type D/IIIB). If we only consider meteors brighter than -3 mag for the calculation of the mean $P_{E}$ value, and meteors fainter than -3 mag for the calculation of the mean $K_B$ value (Table \ref{meanPhysical}), JEO meteoroids are defined by $P_E \approx$ -5.83 (Type IIIB) and $K_B \approx$ 6.97 (Type C1). 

None of the samples exhibited material strengths consistent with ordinary or carbonaceous chondrites. The $P_{E}$ distribution on Fig. \ref{KbPe} also shows that larger JEO meteoroids are generally classified as the most fragile (IIIB). Similar mass dependency was previously revealed also in other stream meteoroids \citep{2017A&A...605A..68S}. 

Figure \ref{KbPe} shows that JEO meteoroids are in terms of material strength comparable to other fragile cometary streams. JEO appear to be slightly stronger than $\alpha$-Capricornids (from comet 169P/NEAT) or $\kappa$-Cygnids. Based on $P_{E}$, they are on average more fragile than Perseids (109P/Swift–Tuttle) and significantly weaker than Taurids (2P/Encke) or Southern $\delta$-Aquariids (96P/Machholz).

\section{Search for the parent object}  \label{Search}

The determined orbital, spectral, and material properties of JEO meteoroids were used as criteria for the identification of the parent object of the stream. In the first step, we identified potential parent objects using a combination of the established orbital similarity D-criteria: the Southworth-Hawkins $D_{SH}$ \citep{1963SCoA....7..261S}, Drummond $D_{D}$ \citep{1981Icar...45..545D} and Jopek criterion $D_{H}$ \citep{1993Icar..106..603J}. The orbital elements of the potential parent objects were taken from the Asteroid Orbital Elements Database\footnote{https://asteroid.lowell.edu/main/astorb}.

Only two bodies were found to satisfy D-criterion threshold values required for primary association ($D_{SH} \leq$ 0.10, $D_{D} \leq$ 0.06, $D_{H} \leq$ 0.10). The mean orbit of the JEO stream and the two closest larger objects (300P and 2009 MU) are given in Table \ref{orbitparent}. Table \ref{orbitparent} also includes the orbit of the impacting 2019 MO which was speculated to be linked to the shower. The second closest body to the stream is small Apollo asteroid 2009 MU (30-75 m in size). The moderate similarity to its orbit is likely only a result of the dense orbital space in this NEO region and, given the much closer association to 300P, it was not further explored.

\begin{table*}[]
\centering
\small\begin{center}
\caption {Comparison of the computed mean orbit of the JEO stream, the proposed parent of the stream 300P/Catalina, dynamically closest asteroid 2009 MU and the impacting asteroid 2019 MO. The orbital similarity criteria (Southworth-Hawkins $D_{SH}$, Drummond $D_{D}$ and Jopek $D_{H}$) are displayed along with specific orbital elements and Tisserand parameter with respect to Jupiter ($T_J$).} 
\vspace{0.1cm}
\begin{tabular}{lrrrrrrrrrl}
\hline\hline\\
\multicolumn{1}{c}{}& %
\multicolumn{1}{c}{$D_{SH}$} & %
\multicolumn{1}{c}{$D_{D}$} & %
\multicolumn{1}{c}{$D_{H}$} & %
\multicolumn{1}{c}{$a$} & %
\multicolumn{1}{c}{$e$} & %
\multicolumn{1}{c}{$q$} & %
\multicolumn{1}{c}{$Q$} & %
\multicolumn{1}{c}{$i$} & %
\multicolumn{1}{c}{$\Pi$}& %
\multicolumn{1}{c}{$T_J$} %
\\
\hline\\
JEO mean      &       &      &      & 2.79  & 0.68 & 0.884  & 4.69 & 5.16 & 319.70 & 2.94 \\
   &              &       &      &  $\pm$ 0.20  & 0.02 & 0.006  & 0.41 & 0.42 & 0.95   & 0.13 \vspace{0.1cm}\\
300P/Catalina & 0.05  & 0.03 & 0.06 & 2.70  & 0.69 & 0.833  & 4.56 & 5.68 & 318.53 & 2.96 \vspace{0.1cm}\\
2009 MU           & 0.09  & 0.06 & 0.08 & 2.29  & 0.61 & 0.891  & 3.68 & 7.29 & 318.80 & 3.32 \vspace{0.1cm}\\
2019 MO       & 0.16  & 0.07 & 0.14 & 2.53  & 0.63 & 0.938  & 4.11 & 1.56 & 307.82 & 3.14 \\
\hline
\end{tabular}
\label{orbitparent}
\end{center}
\end{table*}

The potential connection of the JEO outburst to the impact of small asteroid 2019 MO was first suggested by Denis Denisenko based on the time correlation and similar entry speed. Our results however show that the orbits of 2019 MO and JEO meteoroids are quite distant with $D_{SH} =$ 0.16 (Table \ref{orbitparent}). In fact, in our search we have found 39 other minor bodies with closer orbits. 

Following the impact of 2019 MO in the Earth's atmosphere, radar record signatures of falling meteorites were reported in data from the NEXRAD weather radar network operated by NOAA, the TJUA (San Juan, Puerto Rico). The first appearance of falling meteorites on radar occurred at 21:26:15 UTC and 10.6 km above sea level (report summarized in the NASA's Astromaterials Research \& Exploration Science page\footnote{https://ares.jsc.nasa.gov/meteorite-falls/}). It was estimated that all meteorites from this event ended up on the seafloor at a depth of approximately 4.8 km.

The determined fragile material strengths of JEO meteoroids are not consistent with a material which could produce meteorite fragments. As noted in Section \ref{Structure}, JEO meteoroids are of significantly weaker material strength compared to Taurids (Fig. \ref{KbPe}), which are, according to the latest studies, not likely to be capable of producing meteorites \citep{2017A&A...605A..68S, 2017P&SS..143..104M}. Furthermore, the estimated size of 2019 MO was only around four meters. Fragile cometary body of such size would likely disintegrate completely before reaching the lower atmosphere. This is another piece of evidence supporting the conclusion that 2019 MO and JEO do not share the same origin.

The most consistent link considering both dynamical and physical properties of JEO meteoroids is to comet 300P/Catalina.  The found current orbital similarity between 300P and the mean JEO orbit ($D_{SH}$ = 0.05, $D_{D}$ = 0.03, $D_{H}$ = 0.06) satisfies requirements for close association in all considered D-criteria. JEO meteoroids show typical cometary material strengths and spectra similar to chondritic materials, which can also be seen in other cometary meteoroids. Besides 300P, there are no other known major cometary bodies on orbits similar to the JEO stream. 

300P was only the second comet to ever be listed in the Sentry Risk Table, which contains minor bodies with the highest risk of future Earth impact. It has since been removed but it is still defined by Earth-MOID of 0.0245 au. A close approach of the comet to Earth was however not the cause of the increased JEO meteor activity in 2019. At the time of the outburst (June 21 - 25), the comet's geocentric distance was between 3.74 and 3.78 au. The observed cluster of meteoroids was therefore created during previous perihelion passages. 

With orbital period of 4.4 years, 300P is a comet with the ninth shortest orbital period known. Besides its close approaches to Earth, 300P makes frequent close approaches to other planets. Most notably, at its aphelion it approaches Jupiter (Jupiter MOID = 0.817 au), which likely affects its dynamical stability. Based on the radar observations of \citet{2006Icar..184..285H}, 300P has a small ($\approx$ 1.4 km), rough, and rapidly rotating nucleus. It was estimated that the grains of the comet are relatively large ($>$cm) and they have low escape velocities ($\approx$1 m/s) and a low production rate \citep{2006Icar..184..285H}. These properties are consistent with those observed in JEO meteoroids. The stream is generally low in production but given the activity in 2019, it contains larger fragments. The computed mean JEO orbit is very close to the comet's orbit, which is consistent with the low escape velocities.

\section{Dynamical evolution of JEO and 300P/Catalina}  \label{evolution}

We then focused on investigating the dynamical stability of the stream and its proposed parent comet 300P, which allows us to study the origin of the JEO stream and predict its evolution. First, we performed a backwards integration for 5000 years of the orbit of comet 300P and simulated 18 particles distributed evenly along the computed mean JEO orbit (defined by mean anomaly).

To check the reliability of our integration, we performed several tests. Comet 300P and all massive objects were integrated backward and subsequently back to their initial state. The differences between initial and final values of orbital elements (semi-major axis, eccentricity, inclination, argument of perihelion, longitude of ascending node) were maximally of an order of $10^{-3}$ in the case of a 1500-year integration into the past. For the mean anomaly, the difference was $0.11^{\circ}$. Integrations for a longer period of time showed significant change in mean anomaly. However, differences for orbital elements, apart from the mean anomaly, remained below $\sim 10^{-1}$ even for 2500 year backward integration. These results seem to be satisfactory for our conclusions.

\begin{figure*}[t]
\begin{center}
\includegraphics[width=.33\textwidth]{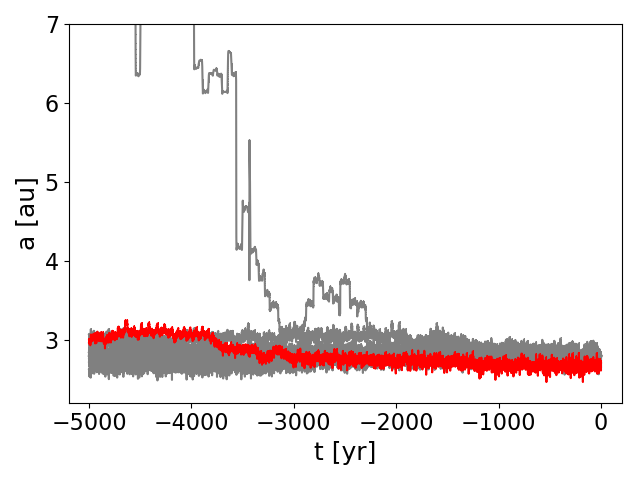}
\includegraphics[width=.33\textwidth]{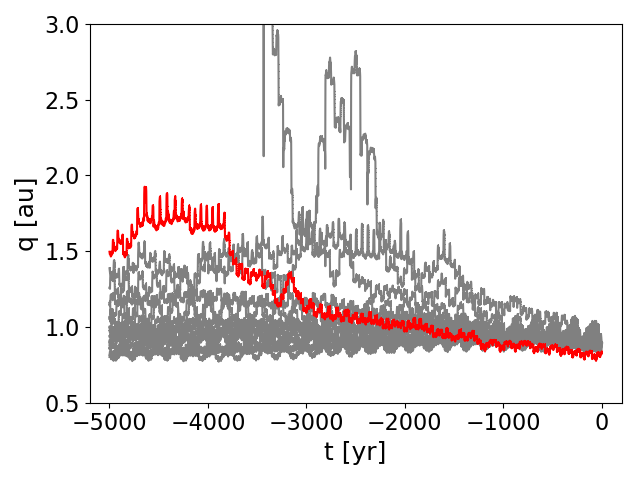}
\includegraphics[width=.33\textwidth]{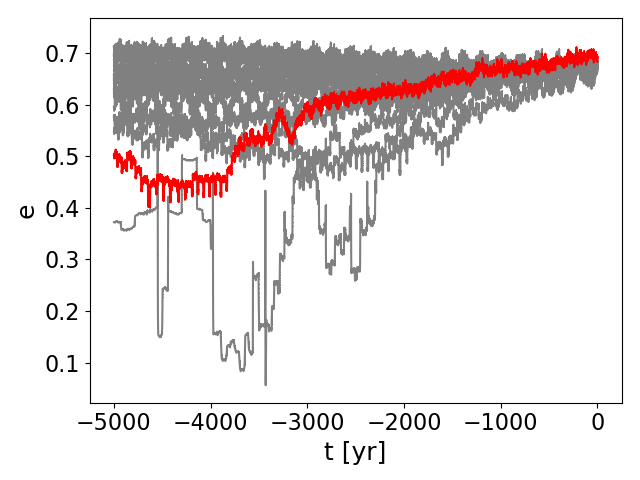}
\includegraphics[width=.33\textwidth]{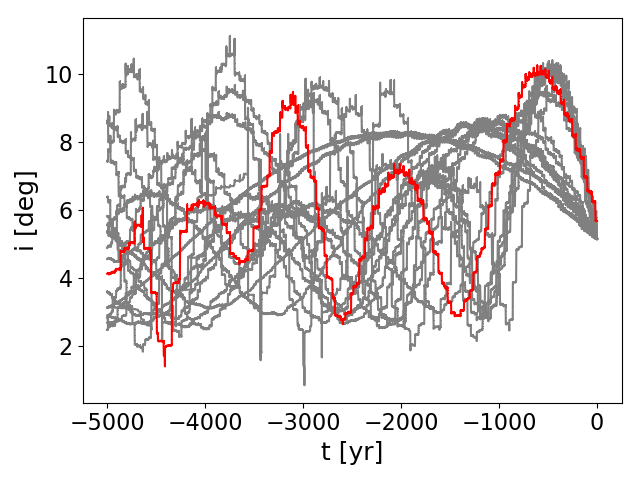}
\includegraphics[width=.33\textwidth]{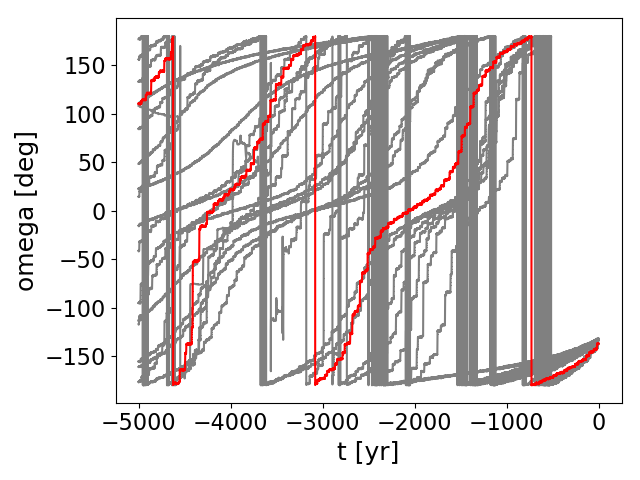}
\includegraphics[width=.33\textwidth]{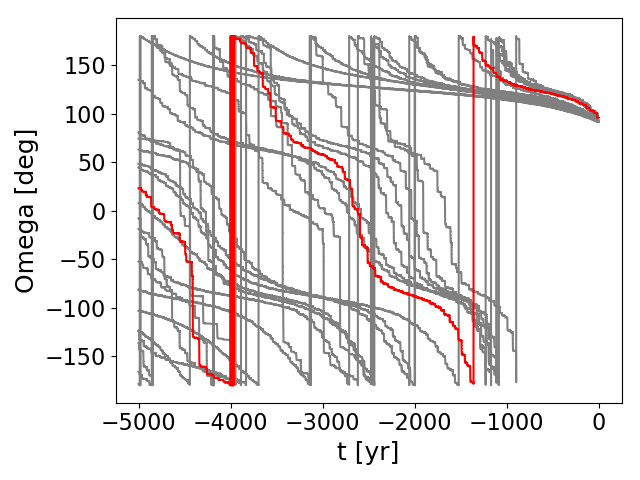}
\includegraphics[width=.33\textwidth]{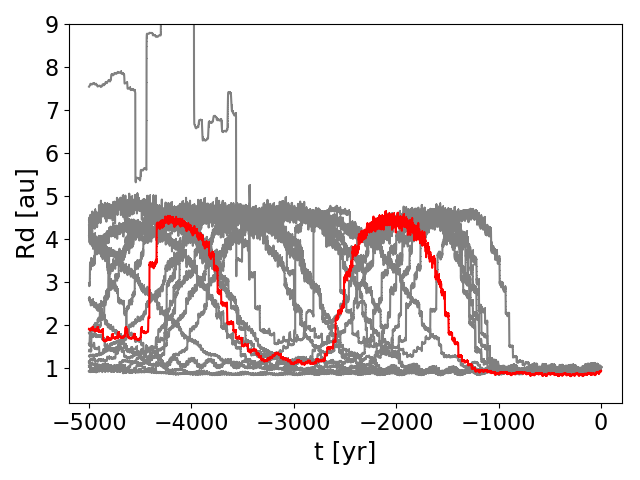}
\includegraphics[width=.33\textwidth]{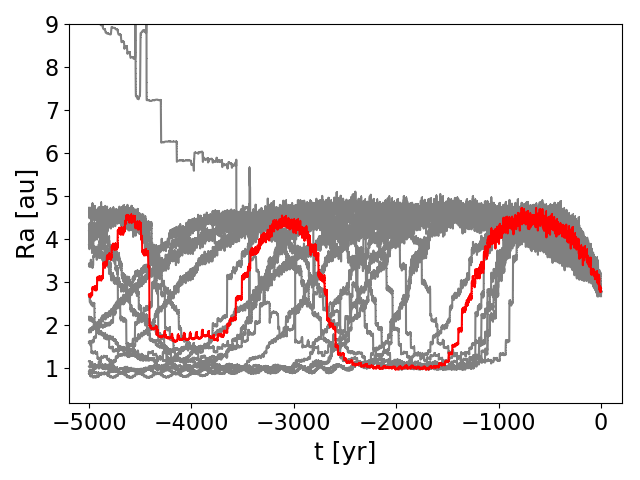}
\end{center}
\caption{Backwards integration of the orbital elements of comet 300P/Catalina (red) and 18 particles representing meteoroids (grey) distributed evenly along the determined mean JEO orbit, defined by different mean anomaly.}%
\label{300Porbit}%
\end{figure*}

The dynamical evolution of comet 300P (Fig. \ref{300Porbit}) reveals significant changes in orbital elements on a relatively short time-scale. Fig. \ref{300Porbit} shows gradual increase of eccentricity from 0.4 to 0.7 and decrease of semi-major axis from 3.1 au to the current 2.7 au in the last 4000 years. We furthermore observe oscillations of the orbit inclination between 2$^{\circ}$ and 10$^{\circ}$ with a period of approximately 1000-1200 years. The instability of the orbit is mainly caused by gravitational perturbation during close aphelion approaches to Jupiter. The simulated particles distributed along the mean JEO orbit quickly deviate from its original orbit. On a timescale of approximately 1000 years, most of the simulated particles move to a orbit which would not be associated with its parent body. 

Interestingly, Fig. \ref{300Porbit} also shows that the perihelion distance $q$ dropped below 1.0 au (condition for the encounter with Earth) only approximately 2000 years ago. This means that particles released from comet 300P before this point could not have been observed as a meteor shower on Earth. The evolution of the heliocentric distance of the descending node (Rd at Fig. \ref{300Porbit}) shows that the comet could have only produced the meteoroid stream approximately in the last 1000 years (when Rd $\approx$ 1 au). 

\begin{figure*}[h]
\begin{center}
\includegraphics[width=.45\textwidth]{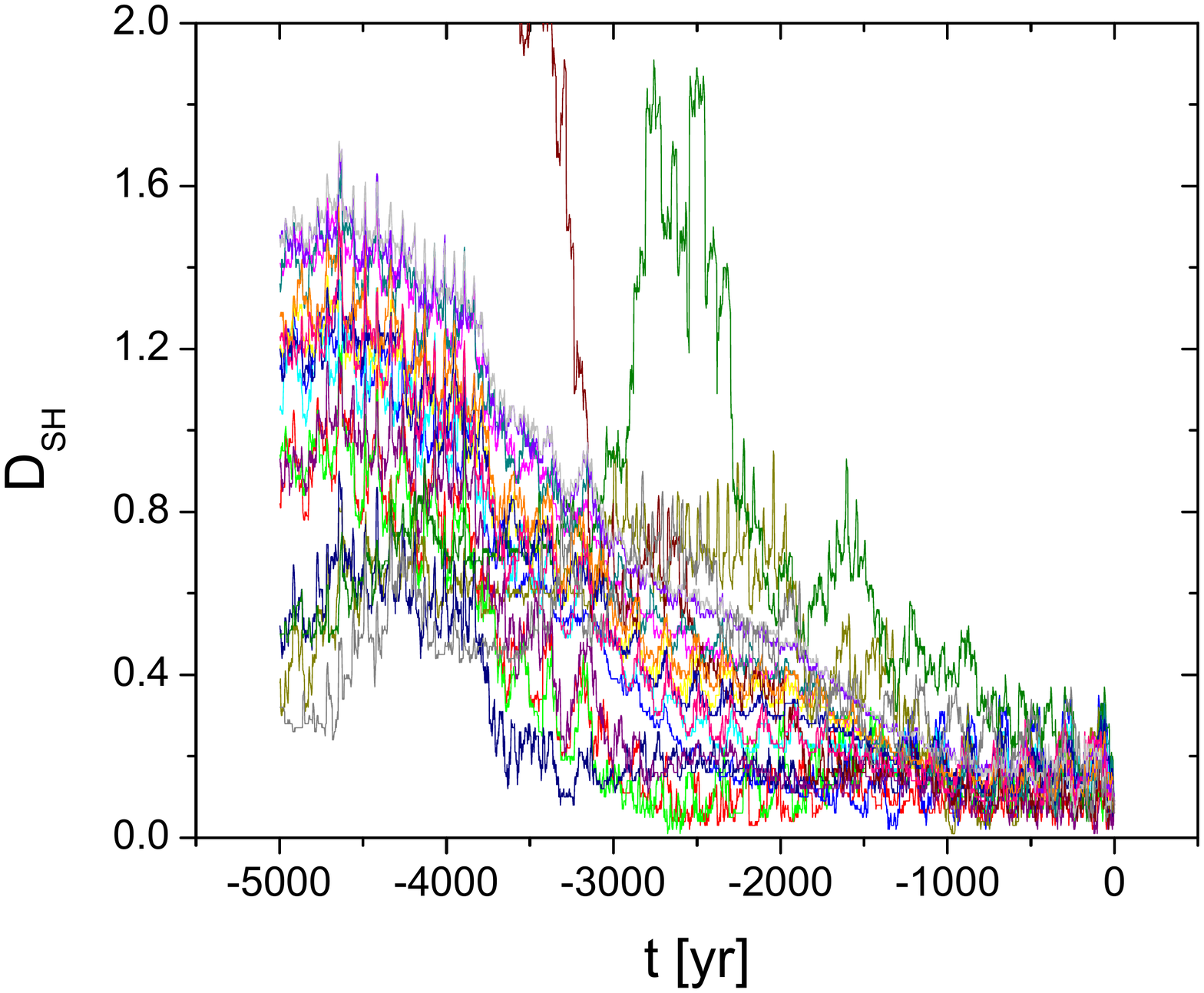}
\includegraphics[width=.45\textwidth]{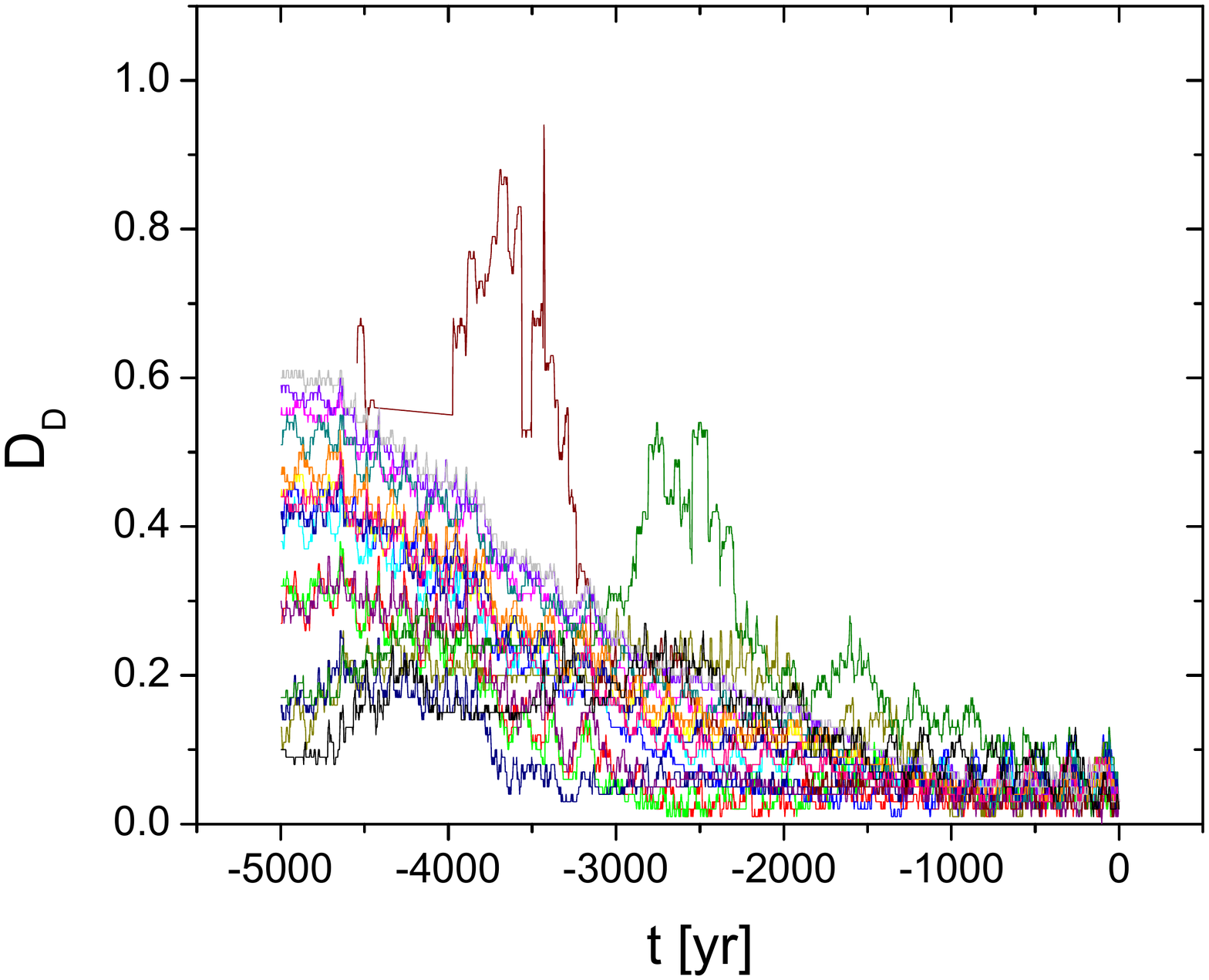}
\end{center}
\caption{Backwards integration of 18 particles positioned along the calculated mean JEO orbit (Table\ref{meanorbit}) and their distance to the orbit of comet 300P/Catalina evaluated based on the $D_{SH}$ and $D_{D}$ criterion.}%
\label{Dint}%
\end{figure*}

Figure \ref{Dint} shows the evolution of the orbital similarity between the orbits of simulated meteoroids and comet 300P. Their distance was evaluated on a scale of 500 years, using the $D_{SH}$ and $D_{D}$ criteria. This allows us to study the stability of the JEO meteoroid stream. The results of this integration (Fig. \ref{Dint}) confirm that JEO meteoroids can only sustain on orbits close to its parent object for a time-scale of up to 1000 years. After this time, most simulated particles moved to orbits that would not be associated with comet 300P. The orbital evolution of meteoroids released from the comet greatly depends  on the location on the orbit where the release occurs.

We also investigated whether 300P comet could be in mean motion resonance (MMR) or Kozai resonance. To study $p : (p + q)$ MMR, we adopted the resonant argument $\varphi = (p + q)\lambda_{300P}-p\lambda_{planet}-q\varpi_{300P}$ , where $\lambda_{300P}$, $\lambda_{planet}$ denote the mean longitudes of 300P and particular planet and $\varpi_{300P}$ represents longitude of pericenter of comet 300P \citep{1999ssd..book.....M}. In the case of MMR, resonant angles oscillate around constant value. To reveal the possibility of being in Kozai resonance, we searched for characteristic behavior (spiraling around a point) in $e - \omega$ phase space and constant value of $\sqrt{1-e^2}\cos i$. According to our integration, 300P was in 8:3 and 5:2 MMR with Jupiter for short period of time approximately 1200 and 3200 years ago, respectively. No evidence of Kozai resonance was found.

\section{Conclusions}

We confirm the unexpected activity of the June epsilon Ophiuchids in 2019 and provide a dynamical, spectral, and physical characterization of the meteoroid stream. The JEO meteoroids are of low material strength that is consistent with fragile cometary structure. The larger meteoroids in our sample were on average classified as type IIIB material, characteristic for very soft (Draconid type) cometary material. The beginning and terminal points of their luminous trajectories are mass-dependent and the light curves exhibit numerous flares, implying a very porous structure for the meteoroids. 

JEO meteors exhibit specific spectral signatures. The spectra show slightly increased iron content compared to other cometary streams and given their speed, a low Na/Mg intensity ratio. They were classified as Na-poor, but the process of Na depletion is not clear. It might trace back to space weathering of the original cometary surface. Besides the most prominent Fe I, Na I, and Mg I multiplets, JEO spectra include lines of Cr I, Ca I, and Mn I. Overall, the spectra are consistent with a primitive chondritic composition, but might reflect compositional difference of their parent comet compared to other stream-producing comets.

Based on the determined set of parameters, we confirm that the JEO stream is most likely to have originated in the short-period comet, 300P/Catalina, which is the only object with a consistent link both in terms of orbital similarity and physical properties of studied meteoroids. Our results suggest that the impact of the small asteroid 2019 MO, on June 22 near Puerto Rico, was not connected to the activity of the JEO outburst.

The backwards integration of the 300P orbit has revealed relatively turbulent dynamical evolution of the comet, affected by close approaches to Earth and Jupiter. We also found that 300P was in 8:3 and 5:2 MMR with Jupiter for short period of time approximately 1200 and 3200 years ago. We estimate that the JEO stream must have been formed within the last 1000 years. Over larger timescales, most of the simulated meteoroids drift to orbits which would not be associated with the comet. The observed JEO outburst points out the past activity of the comet, which could also produce enhanced activity within the meteor shower in the future.

\begin{acknowledgements}

This work was supported by the Slovak Research and Development Agency under the contract APVV-16-0148 and by the Slovak Grant Agency for Science grant VEGA 01/0596/18. We are grateful to the support staff contributing to the operation of the AMOS network in Slovakia and to the Instituto de Astrof\'{i}sica de Canarias for providing support with the installation and maintenance of AMOS systems on the Canary Islands. We are thankful to David Asher for his insightful review.
     
\end{acknowledgements}

\bibliographystyle{aa}
\bibliography{references}

\begin{appendix}

\section{Orbital and trajectory parameters} 

\begin{table*}[h]
\centering
\small\begin{center}
\caption {Trajectory and physical properties of JEO meteoroids. Each meteoroid is designated with characteristic code based on the date and time of observation in UTC (format yyyymmdd\_hhmmss), absolute magnitude (Mag), estimated photometric mass $M_P$, radiant right ascension $RA_g$ in deg, radiant declination $dec_g$ in deg, initial velocity $v_i$ in km\,s\textsuperscript{-1}, beginning height $H_B$, terminal height $P_E$ , and trajectory length $L$ in km, $K_B$ and $P_E$ parameters and corresponding material strength classes.} 
\vspace{0.1cm}
\begin{tabular}{lrrrrrrrrrcrcc}
\hline\hline\\
\multicolumn{1}{c}{Meteor}& %
\multicolumn{1}{c}{Mag}& %
\multicolumn{1}{c}{$M_P$} & %
\multicolumn{1}{c}{$RA_g$} & %
\multicolumn{1}{c}{$dec_g$} & %
\multicolumn{1}{c}{$v_i$} & %
\multicolumn{1}{c}{$H_B$} & %
\multicolumn{1}{c}{$H_E$}& %
\multicolumn{1}{c}{$L$}& %
\multicolumn{1}{c}{$K_B$}& %
\multicolumn{1}{c}{}& %
\multicolumn{1}{c}{$P_E$}& %
\multicolumn{1}{c}{} %
\\
\hline\\
M20190622\_222433 & -5.2 & 230  & 244.28 & -8.47  & 18.66     & 91.95   & 66.59   & 31.3   & 6.99 & C1 & -5.68 & IIIA \\
& $\pm$ 0.6 &  & 0.16 & 0.15 & 0.29 & 0.19 & 0.22 &  & 0.01 &  & 0.11 & \vspace{0.1cm} \\
M20190622\_233116 & -0.5 & 2    & 244.79 & -7.65  & 18.45     & 90.28   & 75.70   & 17.1   & 7.14 & B  & -5.43 & IIIA \\
& $\pm$ 0.7 &  & 0.62 & 0.87 & 1.83 & 1.46 & 1.17 &  & 0.11 &  & 0.21 & \vspace{0.1cm} \\
M20190623\_002905 & 0.8  & 1    & 242.61 & -9.84  & 17.01     & 89.14   & 78.39   & 13.4   & 7.18 & B  & -5.42 & IIIA \\
& $\pm$ 0.8 &  & 0.49 & 0.54 & 0.69 & 0.45 & 0.35 &  & 0.04 &  & 0.11 & \vspace{0.1cm} \\
M20190623\_002925 & -2.5 & 19   & 242.90 & -10.28 & 17.83     & 91.70   & 70.32   & 27.3   & 7.03 & C1 & -5.46 & IIIA \\
& $\pm$ 0.9 &  & 0.18 & 0.17 & 0.05 & 0.28 & 0.21 &  & 0.03 &  & 0.17 & \vspace{0.1cm} \\
M20190623\_011830 & -5.4 & 374  & 245.81 & -9.05  & 18.32     & 100.96  & 71.10   & 39.5   & 6.33 & D  & -6.02 & IIIB \\
& $\pm$ 0.7 &  & 0.08 & 0.10 & 0.09 & 0.08 & 0.09 &  & 0.01 &  & 0.12 & \vspace{0.1cm} \\
M20190623\_012319 & -4.0 & 54   & 245.53 & -7.56  & 18.09     & 91.45   & 71.27   & 20.6   & 7.04 & C1 & -5.84 & IIIB \\
 &$\pm$ 0.5 &  & 0.18 & 0.19 & 0.39 & 0.76 & 0.08 &  & 0.01 &  & 0.08 & \vspace{0.1cm} \\
M20190623\_014851 & -2.0 & 7    & 245.42 & -8.49  & 17.29     & 90.88   & 80.73   & 14.4   & 7.12 & B  & -5.94 & IIIB \\
&$\pm$ 0.6 &  & 0.14 & 0.17 & 0.09 & 0.12 & 0.11 &  & 0.01 &  & 0.11 & \vspace{0.1cm} \\
M20190623\_020529 & -1.0 & 3    & 245.17 & -8.31  & 17.61     & 90.56   & 80.00   & 15.5   & 7.18 & B  & -5.69 & IIIA \\
 &$\pm$ 0.6 &  & 0.11 & 0.10 & 0.15 & 0.13 & 0.11 &  & 0.01 &  & 0.11 & \vspace{0.1cm} \\
M20190623\_214019 & -0.8 & 3    & 244.78 & -9.28  & 17.94     & 94.13   & 80.35   & 18.2   & 6.73 & C1 & -5.78 & IIIB \\
&$\pm$ 0.8 &  & 0.10 & 0.22 & 0.13 & 0.13 & 0.07 &  & 0.02 &  & 0.14 & \vspace{0.1cm} \\
M20190623\_220024 & -2.4 & 17   & 245.04 & -7.79  & 18.36     & 97.93   & 73.66   & 28.8   & 6.41 & D  & -5.69 & IIIA \\
 &$\pm$ 0.7 &  & 0.04 & 0.08 & 0.12 & 0.10 & 0.10 &  & 0.01 &  & 0.13 & \vspace{0.1cm} \\
M20190623\_221410 & -1.8 & 10   & 247.08 & -8.96  & 18.26     & 98.77   & 76.52   & 28.5   & 6.36 & D  & -5.75 & IIIB \\
&$\pm$ 0.9 &  & 0.56 & 0.41 & 0.37 & 0.47 & 0.40 &  & 0.04 &  & 0.18 & \vspace{0.1cm} \\
M20190623\_224636 & -3.0 & 20   & 245.68 & -7.68  & 17.59     & 89.54   & 74.42   & 17.8   & 7.12 & B  & -5.81 & IIIB \\
&$\pm$ 0.6 &  & 0.06 & 0.14 & 0.21 & 0.17 & 0.25 &  & 0.02 &  & 0.12 & \vspace{0.1cm} \\
M20190623\_231128 & 1.1  & 1    & 245.38 & -8.02  & 17.87     & 89.63   & 78.95   & 12.6   & 7.14 & B  & -5.37 & IIIA \\
&$\pm$ 0.8 &  & 0.08 & 0.36 & 0.64 & 0.20 & 0.15 &  & 0.02 &  & 0.14 & \vspace{0.1cm} \\
M20190623\_235555 & 1.3  & 1    & 245.72 & -7.27  & 18.47     & 92.97   & 81.77   & 13.3   & 6.91 & C1 & -5.51 & IIIA \\
&$\pm$ 0.7 &  & 0.06 & 0.14 & 0.26 & 0.08 & 0.08 &  & 0.01 &  & 0.12 & \vspace{0.1cm} \\
M20190624\_001116 & 1.1  & 1    & 245.31 & -8.04  & 17.69     & 89.25   & 77.91   & 13.8   & 7.20 & B  & -5.29 & IIIA \\
 &$\pm$ 0.7 &  & 0.06 & 0.11 & 0.08 & 0.12 & 0.11 &  & 0.01 &  & 0.12 & \vspace{0.1cm} \\
M20190624\_001518 & -2.6 & 20   & 245.51 & -8.28  & 18.12     & 94.44   & 72.35   & 26.9   & 6.81 & C1 & -5.62 & IIIA \\
 &$\pm$ 0.8 &  & 0.08 & 0.16 & 0.26 & 0.09 & 0.09 &  & 0.01 &  & 0.14 & \vspace{0.1cm}\\
M20190624\_003723 & -0.3 & 2    & 245.44 & -7.65  & 18.09     & 89.68   & 72.70   & 20.7   & 7.21 & B  & -5.22 & II   \\
&$\pm$ 0.7 &  & 0.18 & 0.23 & 0.47 & 0.33 & 0.36 &  & 0.03 &  & 0.14 &  \vspace{0.1cm}\\
M20190624\_004801 & -0.9 & 3    & 245.38 & -7.54  & 18.16     & 95.46   & 80.12   & 19.0   & 6.76 & C1 & -5.81 & IIIB \\
&$\pm$ 0.7 &  & 0.17 & 0.21 & 0.40 & 0.12 & 0.07 &  & 0.01 &  & 0.13 &  \vspace{0.1cm}\\
M20190624\_010624 & -1.8 & 7    & 245.62 & -7.95  & 18.32     & 92.89   & 76.86   & 20.9   & 7.00 & C1 & -5.69 & IIIA \\
&$\pm$ 0.6 &  & 0.10 & 0.15 & 0.23 & 0.13 & 0.13 &  & 0.01 &  & 0.11 &  \vspace{0.1cm}\\
M20190624\_034139 & -0.6 & 3    & 245.73 & -8.03  & 17.65     & 93.61   & 81.97   & 26.2   & 7.07 & C1 & -5.60 & IIIA \\
 &$\pm$ 0.8 &  & 0.11 & 0.07 & 0.06 & 0.16 & 0.14 &  & 0.02 &  & 0.14 & \vspace{0.1cm}\\
M20190624\_222248 & -4.1 & 86   & 245.78 & -7.56  & 17.86     & 94.18   & 70.78   & 16.1   & 6.75 & C1 & -5.80 & IIIB \\
&$\pm$ 0.7 &  & 0.04 & 0.10 & 0.07 & 0.21 & 0.15 &  & 0.02 &  & 0.12 &  \vspace{0.1cm}\\
M20190625\_012948 & -1.7 & 8    & 245.31 & -7.45  & 17.62     & 88.54   & 71.49   & 22.5   & 7.31 & A  & -5.38 & IIIA \\
&$\pm$ 0.5 &  & 0.09 & 0.13 & 0.12 & 0.13 & 0.36 &  & 0.01 &  & 0.09 &  \vspace{0.1cm}\\
\hline
\end{tabular}
\label{physical}
\end{center}
\end{table*}

\begin{table*}[t]
\centering
\small\begin{center}
\caption {Orbital properties of JEO meteoroids. Each meteoroid is designated with a code corresponding to the meteor order in Table \ref{physical}, geocentric speed $v_g$, semi-major axis $a$ in au, eccentricity $e$, perihelion distance $q$ an aphelion distance $Q$ in au, inclination $i$, argument of perihelion $\omega$ , and longitude of the ascending node $\Omega$ in deg and Tisserand parameter with respect to Jupiter ($T_J$).} 
\vspace{0.1cm}
\begin{tabular}{rrrrrrrrrl}
\hline\hline\\
\multicolumn{1}{c}{Code}& %
\multicolumn{1}{c}{$v_g$} & %
\multicolumn{1}{c}{$a$} & %
\multicolumn{1}{c}{$e$} & %
\multicolumn{1}{c}{$q$} & %
\multicolumn{1}{c}{$Q$} & %
\multicolumn{1}{c}{$i$} & %
\multicolumn{1}{c}{$\omega$}& %
\multicolumn{1}{c}{$\Omega$}& %
\multicolumn{1}{c}{$T_J$} %
\\
\hline\\
M20190622\_222433 & 14.88 & 3.14 & 0.72 & 0.880 & 5.41 & 5.20 & 227.31 & 90.92 & 2.73 \\
&$\pm$ 0.36 & 0.24 & 0.02 & 0.002 & 0.49 & 0.15 & 0.18 &  & 0.12 \vspace{0.1cm}\\
M20190622\_233116 & 14.77 & 3.02 & 0.71 & 0.879 & 5.16 & 5.54 & 227.58 & 90.97 & 2.79 \\
&$\pm$ 2.23 & 0.56 & 0.13 & 0.015 & 1.42 & 1.00 & 0.83 &  & 0.43 \vspace{0.1cm}\\
M20190623\_002905 & 13.05 & 2.46 & 0.63 & 0.898 & 4.01 & 4.03 & 225.62 & 90.99 & 3.18 \\
&$\pm$ 0.91 & 0.34 & 0.05 & 0.007 & 0.69 & 0.43 & 0.45 &  & 0.28 \vspace{0.1cm}\\
M20190623\_002925 & 14.12 & 2.89 & 0.69 & 0.889 & 4.89 & 4.16 & 226.16 & 91.00 & 2.87 \\
&$\pm$ 0.06 & 0.04 & 0.00 & 0.001 & 0.07 & 0.08 & 0.15 &  & 0.02 \vspace{0.1cm}\\
M20190623\_011830 & 14.80 & 2.90 & 0.70 & 0.872 & 4.92 & 5.10 & 229.13 & 91.03 & 2.86 \\
 &$\pm$ 0.11 & 0.05 & 0.01 & 0.001 & 0.11 & 0.06 & 0.12 &  & 0.03 \vspace{0.1cm}\\
M20190623\_012319 & 14.52 & 2.81 & 0.69 & 0.878 & 4.74 & 5.57 & 228.31 & 91.04 & 2.92 \\
 &$\pm$ 0.48 & 0.29 & 0.03 & 0.004 & 0.57 & 0.24 & 0.19 &  & 0.17 \vspace{0.1cm}\\
M20190623\_014851 & 13.55 & 2.42 & 0.63 & 0.884 & 3.95 & 4.91 & 228.41 & 91.05 & 3.20 \\
 &$\pm$ 0.12 & 0.04 & 0.01 & 0.001 & 0.08 & 0.09 & 0.13 &  & 0.03 \vspace{0.1cm}\\
M20190623\_020529 & 13.98 & 2.60 & 0.66 & 0.882 & 4.32 & 5.08 & 228.07 & 91.07 & 3.06 \\
 &$\pm$ 0.19 & 0.07 & 0.01 & 0.002 & 0.14 & 0.10 & 0.11 &  & 0.05 \vspace{0.1cm}\\
M20190623\_214019 & 13.89 & 2.68 & 0.67 & 0.887 & 4.47 & 4.63 & 227.00 & 91.84 & 3.01 \\
 &$\pm$ 0.16 & 0.08 & 0.01 & 0.001 & 0.15 & 0.11 & 0.08 &  & 0.05 \vspace{0.1cm}\\
M20190623\_220024 & 14.47 & 2.94 & 0.70 & 0.884 & 4.99 & 5.38 & 226.87 & 91.86 & 2.84 \\
 &$\pm$ 0.16 & 0.09 & 0.01 & 0.001 & 0.18 & 0.07 & 0.06 &  & 0.05 \vspace{0.1cm}\\
M20190623\_221410 & 14.34 & 2.64 & 0.67 & 0.873 & 4.40 & 5.09 & 229.60 & 91.87 & 3.03 \\
 &$\pm$ 0.46 & 0.22 & 0.03 & 0.004 & 0.45 & 0.23 & 0.79 &  & 0.16 \vspace{0.1cm}\\
M20190623\_224636 & 13.56 & 2.47 & 0.64 & 0.888 & 4.06 & 5.21 & 227.48 & 91.89 & 3.16 \\
 &$\pm$ 0.26 & 0.13 & 0.02 & 0.002 & 0.25 & 0.13 & 0.06 &  & 0.10 \vspace{0.1cm}\\
M20190623\_231128 & 13.98 & 2.67 & 0.67 & 0.886 & 4.45 & 5.18 & 227.24 & 91.90 & 3.01 \\
 &$\pm$ 0.82 & 0.37 & 0.05 & 0.005 & 0.75 & 0.38 & 0.10 &  & 0.26 \vspace{0.1cm}\\
M20190623\_235555 & 14.83 & 3.06 & 0.71 & 0.880 & 5.24 & 5.76 & 227.42 & 91.94 & 2.77 \\
 &$\pm$ 0.32 & 0.18 & 0.02 & 0.002 & 0.37 & 0.15 & 0.05 &  & 0.10 \vspace{0.1cm}\\
M20190624\_001116 & 13.88 & 2.64 & 0.66 & 0.887 & 4.39 & 5.14 & 227.11 & 91.94 & 3.03 \\
 &$\pm$ 0.11 & 0.04 & 0.01 & 0.001 & 0.08 & 0.06 & 0.07 &  & 0.03 \vspace{0.1cm}\\
M20190624\_001518 & 14.43 & 2.87 & 0.69 & 0.882 & 4.86 & 5.23 & 227.45 & 91.95 & 2.88 \\
 &$\pm$ 0.32 & 0.16 & 0.02 & 0.002 & 0.33 & 0.16 & 0.10 &  & 0.10 \vspace{0.1cm}\\
M20190624\_003723 & 14.43 & 2.88 & 0.69 & 0.883 & 4.88 & 5.46 & 227.15 & 91.96 & 2.87 \\
 &$\pm$ 0.58 & 0.36 & 0.03 & 0.004 & 0.70 & 0.28 & 0.18 &  & 0.21 \vspace{0.1cm}\\
M20190624\_004801 & 14.55 & 2.95 & 0.70 & 0.883 & 5.03 & 5.54 & 227.06 & 91.97 & 2.83 \\
 &$\pm$ 0.50 & 0.27 & 0.03 & 0.004 & 0.55 & 0.25 & 0.17 &  & 0.15 \vspace{0.1cm}\\
M20190624\_010624 & 14.78 & 3.05 & 0.71 & 0.880 & 5.22 & 5.47 & 227.46 & 91.98 & 2.78 \\
 &$\pm$ 0.29 & 0.16 & 0.02 & 0.002 & 0.32 & 0.15 & 0.09 &  & 0.08 \vspace{0.1cm}\\
M20190624\_034139 & 14.18 & 2.74 & 0.68 & 0.884 & 4.59 & 5.27 & 227.45 & 92.08 & 2.96 \\
 &$\pm$ 0.07 & 0.03 & 0.00 & 0.001 & 0.06 & 0.05 & 0.12 &  & 0.02 \vspace{0.1cm}\\
M20190624\_222248 & 13.89 & 2.69 & 0.67 & 0.890 & 4.48 & 5.33 & 226.49 & 92.83 & 3.00 \\
 &$\pm$ 0.08 & 0.04 & 0.00 & 0.001 & 0.08 & 0.05 & 0.03 &  & 0.03 \vspace{0.1cm}\\
M20190625\_012948 & 13.95 & 2.78 & 0.68 & 0.892 & 4.66 & 5.35 & 225.79 & 92.95 & 2.94 \\
 &$\pm$ 0.15 & 0.07 & 0.01 & 0.001 & 0.14 & 0.09 & 0.11 &  & 0.04 \vspace{0.1cm}\\
\hline
\end{tabular}
\label{orbits}
\end{center}
\end{table*}

\end{appendix}
\end{document}